\documentclass{aa} 
\usepackage{graphicx}
\usepackage{amsmath}
\usepackage{natbib}
\usepackage{txfonts}

\bibpunct{(}{)}{;}{a}{}{,}

\begin{document}

\title{Coronal properties of the EQ Peg binary system}

\author{C. Liefke\inst{1} \and J.-U. Ness\inst{2} \and J.~H.~M.~M. Schmitt\inst{1} \and A. Maggio\inst{3}}

\offprints{C. Liefke,\\ \email{cliefke@hs.uni-hamburg.de}}

\institute{Hamburger Sternwarte, Univers\"at Hamburg, Gojenbergsweg 112, 21029 Hamburg, Germany
\and 
Arizona State University, School of Earth and Space Exploration, ASU, P. O. Box 871404,Tempe, AZ 85287, USA
\and
INAF -- Osservatorio Astronomico di Palermo, Piazza del Parlamento 1, 90134 Palermo, Italy}

\date{Received 28 April 2008 / Accepted 20 October 2008}

 
\abstract
{The activity indicators of M dwarfs are distinctly different for
early and late types. The coronae of early M dwarfs display
high X-ray luminosities and temperatures, a pronounced inverse
FIP effect, and frequent flaring to the extent that no quiescent level can be defined in many
cases. For late M dwarfs, fewer
but more violent flares have been observed, and the quiescent
X-ray luminosity is much lower.}
{To probe the relationship between coronal properties with spectral type of active M dwarfs, we analyze the M3.5 and M4.5 components of the EQ~Peg binary system in comparison with other active M dwarfs of spectral types M0.5 to M5.5.}
{We investigate the timing behavior of both components of the EQ~Peg system, reconstruct their differential emission measure, and investigate the coronal abundance ratios based on emission-measure independent line ratios from their \emph{Chandra} HETGS spectra. Finally we test for density variations in different states of activity.}
{The X-ray luminosity of EQ~Peg A (M3.5) is by a factor of 6--10 brighter than that of EQ~Peg B (M4.5). Like most other active M dwarfs, the EQ~Peg system shows an inverse FIP effect. The abundances of both components are consistent within the errors; however, there seems to be a tendency toward the inverse FIP effect being less pronounced in the less active EQ~Peg~B when comparing the quiescent state of the two stars. This trend is supported by our comparison with other M~dwarfs.} 
{As the X-ray luminosity decreases with later spectral type, so do coronal temperatures and flare rate. 
The amplitude of the observed abundance anomalies, i.\,e. the inverse FIP effect, declines; however, clear deviations from solar abundances remain.}


\maketitle

\section{Introduction}
M~dwarfs populate the low-mass and low-temperature end of the main sequence, with masses of approximately 0.6~M$_{\sun}$ and an effective temperature of about 3800~K at a spectral type of M0, down to 0.1~M$_{\sun}$ and 2200~K at spectral type M9. Conspicuous changes in the optical spectra become apparent in the later subclasses as more and more molecules form with decreasing photospheric temperature, and molecular bands start to dominate. Dust formation marks the transition from late M~dwarfs to the brown dwarf regime of L and T~dwarfs.

Somewhat surprisingly, all M dwarfs seem to show at least some activity phenomena, and in particular 
active M dwarfs of spectral types M0--M4 are strong coronal X-ray sources with X-ray luminosities often close to a saturation limit of $\log L_{\rm X}$/$L_{\rm bol} \sim -3.3$ \citep{Fleming_M_dwarfs, Pizzolato_saturated}.
Frequent and strong flaring with increases in X-ray luminosity by more than two orders of magnitude \citep[e.\,g.][]{Favata_EV_Lac, Guedel_Proxima_Cen_1, CN_Leo_onset} is widespread among these stars, with additional variability on all timescales and amplitude levels.
Their emission measure distributions peak around 6--8~MK \citep{Robrade_M_dwarfs}, thus
their average coronal temperatures are much higher than solar \citep[peak
temperature 1--2~MK, see e.\,g.][]{Brosius}. Temperatures in excess of several 10~MK 
are typically measured during larger flare events \citep{Robrade_M_dwarfs}.

On the other hand, late M dwarfs with spectral types M6--M9 are difficult to
detect in X-rays at least during quiescence, yet their ability to produce transient X-ray luminosity enhancements by orders of magnitude
during flares \citep{Rutledge_LP944-20, LHS_2065,
Stelzer_LP412-31} remains. 
Intermediate objects with a spectral type of $\approx$~M5
have lower coronal temperatures than the more active
early M~dwarfs, the permanent state of variability is replaced by periods of
quiescence interrupted by individual flares, and the overall X-ray luminosity starts
to decrease by more than an order of magnitude at around
spectral type M5 \citep{Guedel_Proxima_Cen_1, Fuhrmeister_CN_Leo}.

The abundance patterns observed in stellar coronae also appear to be related 
to changes in the coronal activity level. On the one hand, there is the
so-called FIP effect observed in (comparably) inactive stars like
the Sun \citep{Feldman_Laming} or e.\,g. $\alpha$~Cen
\citep{Raassen}, with the coronal abundances of elements with
low first ionization potential (FIP) enhanced with respect to solar photospheric values. On the other
hand, very active stars tend to show the inverse effect,
i.\,e. elements with low FIP depleted and elements with a high
FIP enhanced. 
\citet{Guedel_FIP_IFIP} and \citet{Sun_in_time}
found that the solar-like FIP effect to first disappear and then
reverse with increasing activity for solar analogs at different ages. This correlation can be
expected to also apply to the sequence of M dwarfs: While the
active early M dwarfs are well-known to show the inverse FIP
effect \citep[see e.\,g.][]{Robrade_M_dwarfs},
\citet{Fuhrmeister_CN_Leo} found the M5.5 star CN~Leo to show
much less pronounced abundance anomalies. Unfortunately, 
high-resolution X-ray spectra for intermediate objects are rare; and due to
their low X-ray luminosity, it is in most cases impossible to
obtain spectra suitable for a reasonable abundance analysis for
late M dwarfs to confirm this trend. 

Stellar coronal X-ray emission is associated with the production of magnetic
fields driven by a dynamo mechanism \citep{Parker_dynamo, 
Moffat_dynamo}, which is thought to be located in the transition layers between
the rigidly rotating radiative core and the outer convection
zone that rotates differentially for solar-like stars. However, models of the
stellar interior predict that M~dwarfs of spectral types later
than $\approx$~M3 become fully convective \citep{Chabrier_Baraffe}, leaving no
shear layers between a radiative core and the convection
zone behind. The solar-like $\alpha\omega$-dynamo thus
does not work for these stars, and no dynamo-induced
X-ray emission would be expected. Yet, the observed decline in
X-ray luminosity around spectral type M5--M6 fits the finding of 
\citet{Mullan_fullyconvective} that the influence of magnetic fields can shift the fully convective cut-off towards later spectral types. 
Fully convective stars produce less quiescent X-ray emission, but
other activity indicators like H$\alpha$ emission even increase in strength
towards the low-mass L dwarfs. Alternative dynamo mechanisms, driving the required magnetic fields, have
been proposed for these objects \citep[e.\,g.][]{Durney_dynamo}.
In these scenarios with $\alpha^2$ or turbulent dynamos, the magnetic fields are sustained by small-scale velocity fields, while large-scale rotation plays only a minor role. 
As a consequence, no activity cycles and magnetic structures of small size and with a more uniform surface distribution are expected. However, most recent non-linear 3D simulations of dynamo action in fully convective stars \citep[e.\,g.][]{Chabrier_Kuker, Dobler_fullyconvective, Browning_fullyconvective} have shown that also large-scale fields could be produced in fast-rotating late M~dwarfs, but these models disagree about whether or not the magnetic fields have axisymmetric components and/or a cyclic behavior.

With the required change of the dominating dynamo mechanism
operating in early and late M~dwarfs in mind, changes in the
abundance patterns between these objects are of particular
interest to characterize the coronae induced by the different
dynamos. The above considerations concerning luminosity
demand investigations of coronal properties of M dwarfs located
close to the expected transition. The binary system
\object{EQ Peg} consists of an M3.5 and M4.5 star and thus
provides the rare opportunity to compare two such stars, which
should otherwise be very similar since they are co-eval. 

In the following, we
present our \emph{Chandra} HETGS observations of EQ~Peg, that
allows us to perform a separate spectral analysis of the two
components. We also compare our observations of
EQ~Peg A and B to HETGS spectra of other M dwarfs covering
a range of spectral types. We propose a classification of the
coronal properties of active M dwarfs based on their spectral type. 

\section{The EQ Peg binary system}
The visual binary system EQ~Peg consists of two M dwarfs at a distance of 6.25~pc with an angular separation of $\approx 6''$, i.\,e. EQ~Peg A with spectral type
M3.5 and an apparent magnitude $V_A = 10.35$ and EQ~Peg B with spectral type
M4.5 and magnitude $V_B = 12.4$.
Both components are well-known (optical) flare stars \citep{Pettersen_flarestars, Lacy_flarestars}.
\citet{Jackson_radio} observed microwave emission during
quiescence, which they attributed to the brighter A component.
\citet{Norton_P_rot_EQ_Peg} find a photometric period of 1.0664 days for the EQ~Peg system from data of the SuperWASP transit survey, however, due to the large aperture size of SuperWASP, this period cannot be clearly assigned to one of the two components of EQ~Peg, nor to two other stars in the field of view.
However, \citet{mdwarf_photospheric} find a rotational velocity of 14~km/s for EQ~Peg~A, while \citet{Delfosse_mdwarfs} analyze only the B~component and find 24.2$\pm$1.4~km/s; both $v \sin i$ values are rather high and thus consistent with a short rotation period.

The EQ~Peg system is a strong X-ray and EUV source with a number of flares recorded, but previous missions were not able to angularly resolve the two stars.
\citet{Pallavicini_EXOSAT} discuss two flares observed with \emph{EXOSAT}. The first one, with an
atypically shaped lightcurve, was observed by
\citet{Haisch_EQ_Peg} in the context of a simultaneous
\emph{EXOSAT} and \emph{IUE} campaign. A second large
amplitude flare was observed by \citet{Pallavicini_EQ_Peg_EXOSAT}. \citet{Katsova_EQ_Peg} observed another large flare with \emph{ROSAT} with simultaneous optical photometry. All of these flares were
attributed to the A component. \citet{EQ_Peg_EUVE} observed
another set of flares on EQ~Peg with \emph{EUVE} and
derived a differential emission measure distribution, based on
line flux measurements, with a peak at about 10~MK, which is,
however, not well constrained.  A first
approach to separate the two components in X-rays was undertaken with \emph{XMM-Newton} \citep{Robrade_EQ_Peg}. Although the two stars show considerable
overlap owing to the instrumental point spread function, it was
possible to attribute about three quarters of the overall X-ray
flux to the A component.  A subsequent detailed spectral
analysis without resolving the binary has been
performed by \citet{Robrade_M_dwarfs}. Our \emph{Chandra} HETGS observations
are the first X-ray observations that allow an unambiguous spectral separation of 
the two binary components.

\section{\emph{Chandra} HETGS observations of EQ Peg A and B}
\subsection{Observations and data analysis}
EQ Peg was observed on four occasions between 2006 November 28$^{\rm th}$ and
December 3$^{\rm rd}$ with a total exposure time of 100~ks with the
\emph{Chandra} High Energy Transmission Grating Spectrometer
(HETGS); a journal of the observations is presented in
Table~\ref{log}.  The HETGS consists of two grating arms, the Medium Energy
Grating (MEG) and the High Energy Grating (HEG), which provide X-ray spectra
in the wavelength range between $\approx 1.5$ and 30~\AA\ with medium (MEG: 0.023~\AA\ FWHM) and high (HEG: 0.012~\AA\ FWHM) resolution.\footnote{For a detailed
description of the \emph{Chandra} HETGS see the Chandra Proposers' Observatory Guide at http://asc.harvard.edu/proposer/POG/html/}  For data reduction
we used the CIAO software version 3.4.


\begin{table}
\begin{center}
\caption{\label{log}Observation log for EQ~Peg.}
\begin{tabular}{cccc}
\hline\hline
ObsID& observation time & exp. time [ ks ]\\
\hline
8453 & 2006-11-28T01:51--2006-11-28T22:48 & 29.95\\
8484 & 2006-11-29T16:33--2006-11-29T08:59 & 10.95\\
8485 & 2006-11-30T16:59--2006-12-01T08:54 & 30.94\\
8486 & 2006-12-03T07:55--2006-12-03T09:02 & 27.95\\
\hline
\end{tabular}
\end{center}
\end{table}

The separation of the 0$^{\rm th}$ order images of EQ~Peg A
and B is
5.8$''$ in the ACIS-S image, and therefore the dispersed spectra of the
two stars can be easily separated with only a small overlap of
the HEG and MEG grating arms at the shortest wavelengths ($< 1.5$~\AA). We extracted
$1^{\rm st}$ order HEG and MEG spectra for each component from each
of the four datasets using the standard CIAO tools to obtain
grating spectra for multiple sources. 

For our final analysis we created
separate HEG and MEG spectra for EQ Peg A and B, where we combined all
four observations and the plus- and minus dispersion orders. Only for EQ
Peg A have we also inspected the spectra from the individual observations.

For our final analysis, we created separate HEG and MEG spectra for EQ Peg A and B, where we combined all four observations and the plus and minus dispersion orders. Only for EQ
Peg A we have also inspected the spectra from the individual observations.
All line fluxes were measured with our
CORA program \citep{CORA}, fitting the lines with
Moffat line profiles given by
\begin{equation}
I(\lambda) = I_{max}\left(1+\left(\frac{\lambda - \lambda_0}{\Delta \lambda}\right)^2\right)^{-\beta}
\end{equation}
i.\,e., modified Lorentzians, with an exponent
$\beta = 2.4$ and a fixed line width of 0.02~\AA\ for MEG and 0.01~\AA\ for HEG spectra.

\subsection{Lightcurves}
\begin{figure*}
\resizebox{\hsize}{!}{\includegraphics{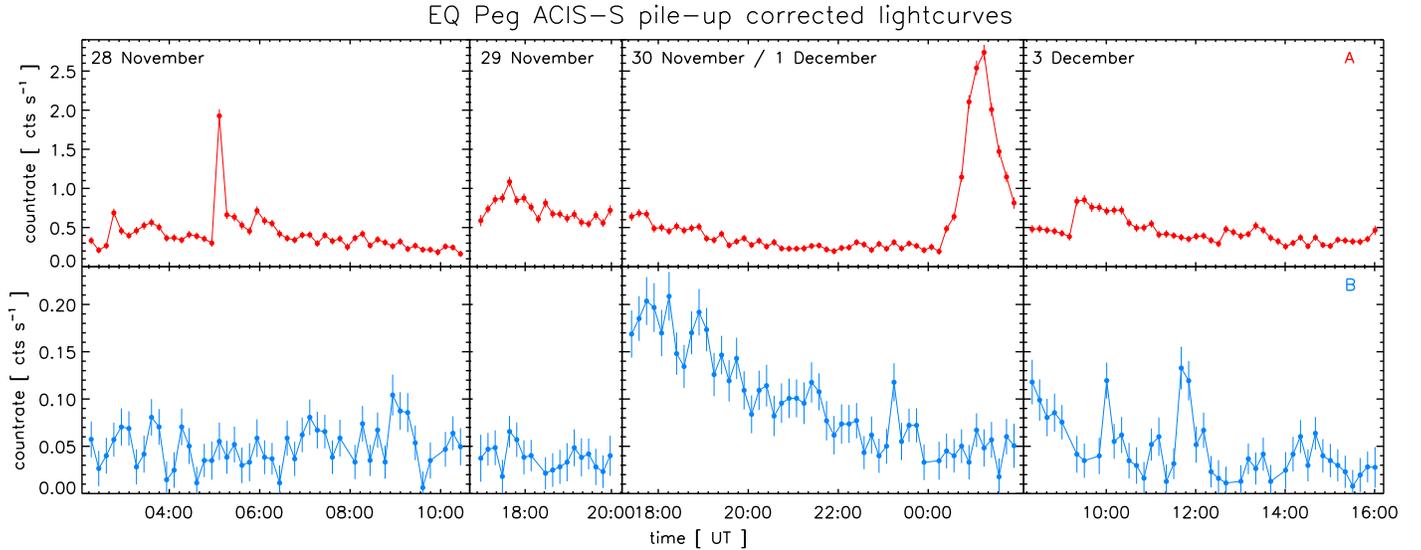}}
\caption[]{\label{lightcurve}Background-subtracted and pileup-corrected lightcurves for EQ~Peg A (top) and B (bottom). Binsize for both sources is 600~s.}
\end{figure*}
We first extracted background-subtracted lightcurves for EQ~Peg A and
B from the dispersed and non-dispersed (0$^{\rm th}$ order)
photons. 
In order to obtain equivalent lightcurves of the two stars with maximized signal-to-noise, we use total (i.\,e. 0$^{\rm th}$ order + dispersed) photons of EQ~Peg~B and the dispersed photons of EQ~Peg~A, scaled by the ratio of total and dispersed photons from EQ~Peg~B (assuming a similar spectral energy distribution for the two stars), as the the 0$^{\rm th}$ order of EQ Peg A is strongly affected by pileup.
The resulting lightcurves are shown in Fig.~\ref{lightcurve} for the four observation intervals.

EQ~Peg~A shows several flares. The most conspicuous event started
at 0:00~UT on December $1^{\rm st}$ with a relatively slow rise and lasted about 2 hours.
Unfortunately, the flare decay phase was not observed completely, since the
observation ended before quiescence was reached again.
At flare peak, the flux increases by a factor of six, and the flare
is strong enough to allow the extraction of separate MEG spectra.
Additional smaller flares occurred on November 28$^{\rm th}$ at
5:00~UT, with a flux increase of about a factor of four, lasting
about half an hour, and on December $3^{\rm rd}$ at about 9:00~UT,
with an increase by a factor of 2, and a longer decay of more
than 2 hours. The underlying quiescent level is difficult to
determine owing to a large number of small-scale variations and
apparently different quiescent emission levels during the four
individual observations, with the base level for the A
component at $\sim 0.3$--0.4 cts/s.  
Apart from pure statistical noise, such low-level amplitude variations can be caused either by small-scale flares or the evolution of active regions on the stellar surface.
On November 29$^{\rm th}$ 
the emission level was much higher than this quiescent level despite
a lack of clear flare-like signatures, which might be explained by active regions newly emerged or rotated into the field of view, while on
November 30$^{\rm th}$ a slow decrease in count rate is observed 
that might be attributed to a long-duration flare. 
The B component at $\sim 0.05$ cts/s is much fainter and does not show conspicuous
short duration flares, except on Dec $3^{\rm rd}$.
On November 30$^{\rm th}$ the count rate of EQ~Peg~B slowly
decreased during the whole 8 hours of the observation and again we may have witnessed the
decay of a flare.

We computed median and mean count rates in order to compare the flux levels of the two components outside obvious flaring periods, and find the quiescence level of EQ~Peg~B lower than
that of A by factors of about 6 to 10.
Both stars show a decay in count rate on November 30$^{\rm th}$, however,
a direct connection of flare activity on the two components is difficult to envisage
considering the distance of $\sim 36$~AU between A and B, 
although a similar correlated behavior of the system with
simultaneous quiescent and flaring states has been observed
by \citet{Robrade_EQ_Peg} with \emph{XMM-Newton}. These authors also found 
a lower flux ratio of 4--5 during quiescence and 
2.5--3 during flaring states.
However, due to the limited spatial resolution of \emph{XMM}'s
EPIC detectors, the exact brightness ratios are less certain.

We also extracted the lightcurves in different energy bands to obtain
time-resolved hardness ratios. 
These give us the temperature
evolution of the plasma. While there is no unique temperature that
describes the spectrum (see Sect.~\ref{DEM_sect}), the temperature distribution
cannot be mapped as a function of time, and the differential emission
measure distribution can only by determined as an average over the entire
observation.
Because of pileup in the
0$^{\rm th}$ order of the A component, we extracted the
hardness ratios only from the dispersed photons for both stars.
With the hard and soft bands chosen to range from 1.0--10.0~keV and 0.3--1.0~keV respectively, the two
components show consistent hardness ratios during quiescence, with values ranging from 0.25 to 0.5, indicating a similar spectral energy distribution. Averaged over the total observing time, the hardness ratio for both stars is 0.43; a spectral hardening to values up to 0.8 is apparent during the major flares on the A component. 
For our spectral analysis, we considered only the total spectrum of EQ~Peg~B, while for EQ~Peg~A we considered the total spectrum and a ``quiescent'' spectrum, where we excluded the two larger flares on November 28$^{\rm th}$ and December $1^{\rm st}$.

We computed overall X-ray luminosities $\log L_{\mathrm{X}}$ of 28.71 for EQ~Peg~A (28.67 for the quiescent state only) and 27.89 for EQ~Peg~B from the integrated dispersed photons in the MEG spectra over a wavelength range from 2 to 25~\AA, i.\,e. in this spectral range, the A component is by a factor of 6--7 more luminous than the B component. However, following \citet{Kenyon_Hartmann}, the bolometric luminosities of the two stars are $\log L_{\mathrm{bol}} \approx 31.90$ and 31.28. This results in $\log L_{\mathrm{X}}/L_{\mathrm{bol}}$ values of $-3.19$ ($-3.23$ during quiescence) and $-3.39$ for EQ~Peg~A and B respectively, i.\,e. both stars are located in the saturation regime of X-ray emission \citep{Pizzolato_saturated}.

\subsection{Grating spectra}
\label{grating}
\begin{figure}
\resizebox{\hsize}{!}{\includegraphics{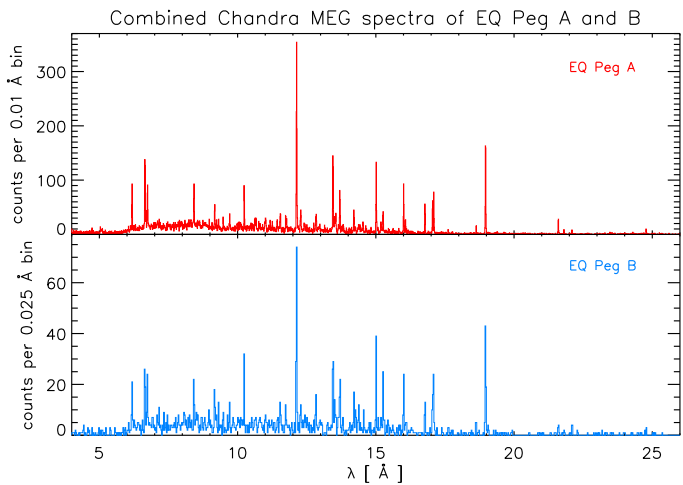}}
\caption[]{\label{spectra}\emph{Chandra} MEG spectra of EQ~Peg A (top, red) and B (bottom, blue). Both spectra have been rebinned.}
\end{figure}

In Fig.~\ref{spectra} we show the \emph{Chandra} MEG spectra of EQ~Peg A
(top) and B (bottom). Both stars show well-pronounced continua between 5 \AA\ and 15 \AA,
indicating larger amounts of high-temperature plasma in their
coronae. The strongest lines in the total MEG count spectra are
\ion{Ne}{x} and \ion{O}{viii} Ly~$\alpha$, other prominent lines
originate from \ion{Si}{xiv} and {\sc xiii}, \ion{Ne}{ix}, and
\ion{Fe}{xvii}. The iron lines tend to be somewhat stronger in
EQ~Peg~B.  The \ion{O}{vii} triplet lines are clearly detected but suffer from
the low effective area of the MEG at long wavelengths. 
The weak \ion{O}{vii} 1s-3p line at 18.63~\AA\ is also detected
in both stars.
After conversion of counts to photon fluxes, the
\ion{O}{viii} Ly~$\alpha$ line is the strongest line in the
spectra, and also the \ion{O}{vii} flux is quite high, indicating
the presence of considerable amounts of low-temperature plasma.

\begin{figure}
\resizebox{\hsize}{!}{\includegraphics{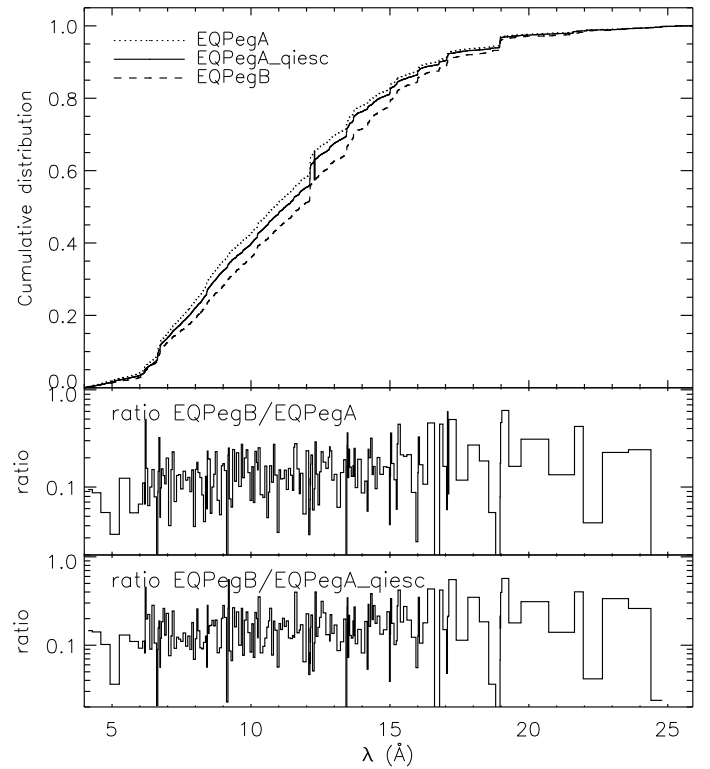}}
\caption[]{\label{cumulative}Cumulative distribution (top) of counts in the MEG spectra of EQ~Peg A (dotted line: all data, solid line: with the large flare on December $1^{st}$ excluded) and B (dashed line); and ratio B\,/\,A (middle panel: all data, bottom: with the two larger flares on EQ~Peg A excluded).}
\end{figure}

In order to probe the shape of the continua and to directly compare the two binary 
components, we
computed the cumulative count spectra from 4~\AA\ to 26~\AA\ from
the combined count spectra of EQ~Peg A and B (top panel of
Figure~\ref{cumulative}). A steep rise of the cumulative spectrum
indicates a harder spectrum and thus higher temperatures, as
observed for EQ~Peg~A compared to EQ~Peg~B. Since this may
reflect the strong flares observed on the A component, we
considered also the quiescent spectrum for EQ~Peg A.
As can be seen from
Figure~\ref{cumulative} (top), this spectrum is indeed closer to
the B component, however, EQ~Peg~A still appears somewhat
harder. When comparing the total spectra of A and B, these
differences are relatively small, and a KS test returns
probabilities of around 3\% that the two stars are different in
their spectral shape. We note that the
presence of strong emission lines, e.\,g., \ion{Ne}{x}
at 12.12~\AA, leads to steps in the
cumulative spectrum. However, the depths of these steps seem
similar for EQ~Peg~A and B, and the major differences in
the cumulative distribution can be attributed to the continuum.

\begin{figure}
\resizebox{\hsize}{!}{\includegraphics{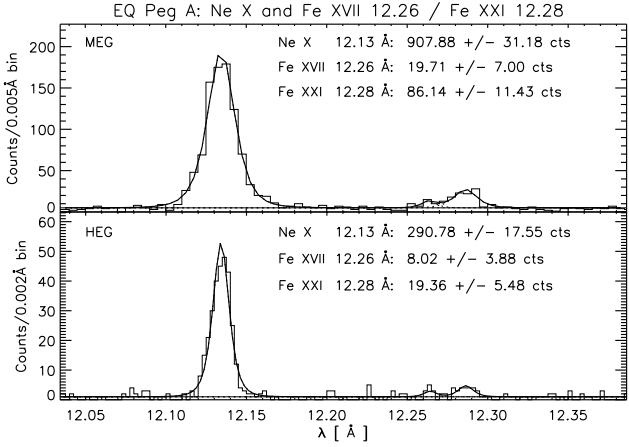}}
\caption[]{\label{cora}Spectral region around the \ion{Ne}{x} line at 12.13~\AA\ and the \ion{Fe}{xvii} and \ion{Fe}{xxi} lines at 12.26 and 12.28~\AA\ in the total spectrum of EQ~Peg~A.}
\end{figure}

In Tables~\ref{counts} and \ref{counts_quiet} we provide lists of the measured line counts and resulting photon fluxes for the combined spectra of EQ¯Peg~A and B, and for the quiescent state of EQ~Peg~A, which we used in our subsequent
analysis; note that a sophisticated procedure has been applied to deconvolve the
blending of the \ion{Ne}{ix} triplet and its contamination with iron lines, mainly from \ion{Fe}{xvii} and \ion{Fe}{xix} \citep{Capella_Ne}. In addition to the line fluxes of the three triplet constituents, we fitted global scaling factors to the line fluxes of the contamining ions observed by \citet{Capella_Ne} in the well-exposed HETGS spectrum of Capella where the contamination is very strong. 
Additionally, the \ion{Ne}{x} line at 12.13~\AA\ has been corrected to account for a blend by \ion{Fe}{xvii} at 12.12~\AA. According to CHIANTI 5.2 \citep{Chianti7}, using the ionization balance from \citet{Mazzotta_etal}, the line emissivity of the contaminating line is comparable to the line at 12.26~\AA, and the ratio of these two lines shows no dependence on electron density and only a mild dependence on electron temperature. Additionally, both transitions go
to the ground state and share the same electron configuration in their excited states. We thus subtracted the line counts measured in the line at 12.26~\AA\ from the \ion{Ne}{x} line as a proxy for the unresolved blending line at 12.12~\AA. Fig.~\ref{cora} illustrates the fit for the total HEG and MEG spectra of EQ~Peg~A.

\begin{table*}
\begin{center}
\caption{\label{counts}Measured line counts and photon fluxes (in $10^{-5}$ photons~cm$^{-2}$ s$^{-1}$) in the combined \emph{Chandra} MEG spectra. Flux predictions from the reconstructed DEMs (see Setc.~\ref{DEM_sect}) are also listed.} 
\begin{tabular}{rrrrrrrrrrrrr}
\hline\hline
$\lambda$& ion  & \multicolumn{4}{c}{EQ Peg A} 				&\multicolumn{4}{c}{EQ Peg B}\\
$[$ \AA\ $]$&	&counts	& photon fluxes	&method 1	&method 2		&counts	& photon fluxes	&method 1	&method 2		 \\
\hline
4.73     &\ion{S}{xvi}		&22.3$\pm$5.6	&0.61$\pm$0.15 	&0.42	&0.32	&---    	&---		&---	&---\\
5.04     &\ion{S}{xv}		&28.7$\pm$6.3	&0.84$\pm$0.18	&0.92	&0.94	&---    	&---		&---	&---\\
6.18     &\ion{Si}{xiv}		&176.6$\pm$14.6 &1.78$\pm$0.15	&1.92	&1.86	&30.4$\pm$5.9   &0.31$\pm$0.06	&0.21	&0.13\\
6.65     &\ion{Si}{xiii}	&321.6$\pm$19.7 &2.74$\pm$0.17	&2.71	&2.75	&40.4$\pm$6.8   &0.34$\pm$0.06	&0.36	&0.37\\
6.69     &\ion{Si}{xiii}	&61.7$\pm$10.6	&0.53$\pm$0.09	&0.72 	&0.73	&5.0$\pm$3.4    &0.04$\pm$0.03	&0.10	&0.10\\
6.74     &\ion{Si}{xiii}	&202.5$\pm$16.2 &1.58$\pm$0.13	&1.31	&1.33	&25.7$\pm$6.0 	&0.20$\pm$0.05	&0.19	&0.19\\
8.42     &\ion{Mg}{xii}		&178.9$\pm$15.9 &1.45$\pm$0.13	&1.58	&1.60	&30.2$\pm$6.0  	&0.24$\pm$0.05	&0.25	&0.23\\
9.17     &\ion{Mg}{xi}		&107.7$\pm$12.4 &1.58$\pm$0.18	&1.54	&1.50	&26.0$\pm$5.6   &0.38$\pm$0.08	&0.32	&0.34\\
9.23     &\ion{Mg}{xi}		&33.9$\pm$8.2	&0.51$\pm$0.12	&0.36	&0.35	&6.7$\pm$3.4   	&0.10$\pm$0.05	&0.08	&0.08\\
9.31     &\ion{Mg}{xi}		&34.6$\pm$8.4	&0.55$\pm$0.13	&0.81	&0.78	&11.2$\pm$3.9  	&0.18$\pm$0.07	&0.17	&0.18\\
10.24    &\ion{Ne}{x}		&160.8$\pm$14.4	&2.42$\pm$0.22	&3.05	&3.07	&29.4$\pm$5.91	&0.44$\pm$0.09	&0.34	&0.36\\
11.17    &\ion{Fe}{xxiv}	&47.2$\pm$9.7	&0.95$\pm$0.19	&0.48	&0.44	&---          	&---		&0.02	&0.01\\
11.55    &\ion{Ne}{ix}		&101.3$\pm$12.3	&2.13$\pm$0.26	&1.80	&1.84	&19.5$\pm$5.12  &0.41$\pm$0.11	&0.28	&0.28\\
11.74    &\ion{Fe}{xxiii}	&65.1$\pm$10.5 	&1.50$\pm$0.24	&0.91	&1.00	&---           	&---		&0.13	&0.09\\
11.78    &\ion{Fe}{xxii}	&44.9$\pm$9.6	&1.11$\pm$0.24 	&1.12	&1.29	&---            &--- 		&0.11	&0.17\\
12.13    &\ion{Ne}{x}		&891.8$\pm$29.9	&24.47$\pm$0.82	&24.00	&24.06	&104.8$\pm$10.6 &2.88$\pm$0.29	&2.75	&2.92\\
13.45    &\ion{Ne}{ix}		&320.2$\pm$17.9	&15.89$\pm$0.88	&17.02	&16.58	&51.4$\pm$7.5	&2.26$\pm$0.33	&2.66	&2.56\\
13.55    &\ion{Ne}{ix}		&77.2$\pm$8.8	&3.28$\pm$0.37	&3.67	&3.57	&9.1$\pm$3.0 	&0.39$\pm$0.13	&0.56	&0.56\\
13.69    &\ion{Ne}{ix}		&184.0$\pm$13.6 &8.90$\pm$0.66	&9.53	&9.28	&32.0$\pm$5.7   &1.55$\pm$0.28	&1.45	&1.46\\
14.21    &\ion{Fe}{xviii}	&92.2$\pm$11.1 	&5.80$\pm$0.70	&4.77	&4.86	&18.7$\pm$4.6	&1.18$\pm$0.29	&0.70	&0.76\\
15.01    &\ion{Fe}{xvii}	&270.1$\pm$17.3 &15.72$\pm$1.01	&18.74	&18.51	&43.2$\pm$7.0	&2.51$\pm$0.40	&3.29	&3.31\\
15.26    &\ion{Fe}{xvii}	&102.2$\pm$9.3	&6.60$\pm$0.60	&5.36	&5.29	&26.5$\pm$5.7   &1.71$\pm$0.36	&0.95	&0.95\\
16.01    &\ion{O}{viii}		&193.1$\pm$14.5 &13.71$\pm$1.03	&8.40	&8.33	&27.2$\pm$5.7	&1.93$\pm$0.40	&0.89	&1.03\\
16.78    &\ion{Fe}{xvii}	&116.4$\pm$11.3 &10.36$\pm$1.00	&11.41	&11.25	&18.3$\pm$4.6	&1.63$\pm$0.41	&2.05	&2.04\\
17.05    &\ion{Fe}{xvii}	&144.7$\pm$12.7 &14.36$\pm$1.26	&14.51	&14.30	&24.8$\pm$5.2   &2.46$\pm$0.52	&2.60	&2.59\\
17.10    &\ion{Fe}{xvii}	&139.3$\pm$12.4 &14.07$\pm$1.25	&12.10	&11.93	&26.2$\pm$5.4 	&2.65$\pm$0.55	&2.15	&2.16\\
18.63    &\ion{O}{vii}		&33.9$\pm$6.2	&4.78$\pm$0.87 	&3.37	&3.43	&7.3$\pm$3.1    &1.03$\pm$0.43	&0.51	&0.48\\
18.97    &\ion{O}{viii}		&435.8$\pm$21.1 &71.90$\pm$3.48	&70.50	&69.90	&63.4$\pm$8.2	&10.46$\pm$1.36	&7.77	&8.82\\
21.60    &\ion{O}{vii}		&69.6$\pm$8.6	&24.21$\pm$2.98	&30.53	&31.30	&7.6$\pm$2.8 	&1.56$\pm$0.91	&4.78	&4.40\\
21.81    &\ion{O}{vii}		&14.2$\pm$4.0	&5.64$\pm$1.59 	&6.83	&7.07	&3.7$\pm$2.0    &1.47$\pm$0.79	&1.45	&0.99\\
22.10    &\ion{O}{vii}		&24.0$\pm$5.1	&10.95$\pm$2.33	&18.08	&18.70	&10.3$\pm$3.5 	&4.70$\pm$1.6	&3.71	&2.62\\
24.79    &\ion{N}{vii}		&19.3$\pm$4.5	&7.84$\pm$1.85 	&7.84	&7.84	&2.0$\pm$1.7    &0.80$\pm$0.72	&0.80	&0.80\\
\hline                                                       
\end{tabular}
\end{center}
\end{table*}

\begin{table}
\begin{center}
\caption{\label{counts_quiet}Measured line counts and photon fluxes in the \emph{Chandra} MEG spectra of the quiescent state of EQ~Peg~A.} 
\begin{tabular}{rrrr}
\hline\hline
$\lambda$ [ \AA\ ] & ion 	&counts		& fluxes	 \\
\hline
4.73     &\ion{S}{xvi}		&8.6$\pm$4.1      &0.25$\pm$0.12 \\
5.04     &\ion{S}{xv}		&22.6$\pm$5.8     &0.70$\pm$0.18 \\
6.18     &\ion{Si}{xiv}		&143.7$\pm$13.2   &1.54$\pm$0.14 \\
6.65     &\ion{Si}{xiii}	&317.7$\pm$19.1   &2.87$\pm$0.17  \\
6.69     &\ion{Si}{xiii}	&57.0$\pm$10.1    &0.52$\pm$0.09  \\
6.74     &\ion{Si}{xiii}	&193.6$\pm$15.6   &1.60$\pm$0.13 \\
8.42     &\ion{Mg}{xii}		&175.4$\pm$15.3   &1.51$\pm$0.13\\	
9.17     &\ion{Mg}{xi}		&98.7$\pm$11.8    &1.54$\pm$0.18 \\
9.23     &\ion{Mg}{xi}		&27.5$\pm$8.1     &0.44$\pm$0.13 \\
9.31     &\ion{Mg}{xi}		&31.9$\pm$8.4     &0.54$\pm$0.14 \\
10.24    &\ion{Ne}{x}		&155.6$\pm$14.0   &2.49$\pm$0.22 \\	
11.17    &\ion{Fe}{xxiv}	&20.6$\pm$7.6     &0.44$\pm$0.16 \\ 
11.55    &\ion{Ne}{ix}		&90.3$\pm$11.5    &2.02$\pm$0.26 \\
11.74    &\ion{Fe}{xxiii}	&45.9$\pm$9.0     &1.12$\pm$0.22 \\ 	
11.78    &\ion{Fe}{xxii}	&42.4$\pm$9.0     &1.11$\pm$0.24\\	
12.13    &\ion{Ne}{x}		&813.8$\pm$28.5   &23.75$\pm$0.83 \\
13.45    &\ion{Ne}{ix}		&293.4$\pm$17.1   &13.67$\pm$0.80 \\	
13.55    &\ion{Ne}{ix}		&61.1$\pm$7.8      &2.76$\pm$0.35 \\ 	
13.69    &\ion{Ne}{ix}		&169.1$\pm$13.0   &8.69$\pm$0.67\\	
14.21    &\ion{Fe}{xviii}	&82.7$\pm$10.6    &5.53$\pm$0.71 \\	
15.01    &\ion{Fe}{xvii}	&266.3$\pm$16.9   &16.45$\pm$1.05 \\	
15.26    &\ion{Fe}{xvii}	&92.8$\pm$10.7    &7.39$\pm$0.77\\	
16.01    &\ion{O}{viii}		&184.8$\pm$14.0   &13.69$\pm$1.05\\	
16.78    &\ion{Fe}{xvii}	&114.3$\pm$11.0   &10.80$\pm$1.04 \\	
17.05    &\ion{Fe}{xvii}	&139.9$\pm$12.6   &14.73$\pm$1.33 \\
17.10    &\ion{Fe}{xvii}	&137.2$\pm$12.4   &14.70$\pm$1.33 \\
18.63    &\ion{O}{vii}		&30.3$\pm$5.6      &4.55$\pm$0.87\\	
18.97    &\ion{O}{viii}		&414.0$\pm$20.5   &72.49$\pm$3.60\\	
21.60    &\ion{O}{vii}		&64.5$\pm$8.3      &23.76$\pm$3.06\\	
21.81    &\ion{O}{vii}		&14.0$\pm$4.0      &5.91$\pm$1.69  \\
22.10    &\ion{O}{vii}		&24.0$\pm$5.1      &11.62$\pm$2.47\\	
24.79    &\ion{N}{vii}		&15.0$\pm$4.3      &6.47$\pm$1.85  \\
\hline                                                       
\end{tabular}
\end{center}
\end{table}

\section{Results}
\subsection{Coronal densities}
\label{dens}
Using the forbidden and intercombination lines of the helium-like triplets of silicon,
magnesium, neon, and oxygen, we computed the density-sensitive f/i-ratios of these
ions for EQ~Peg~A and B. In Table~\ref{densities} we
list these ratios with the respective peak formation temperatures
of the ion and the derived electron densities $n_e$ for EQ~Peg A and B.
For the conversion of the measured 
$f/i$~ratios to densities, we used the relation 
\begin{equation}
\frac{f}{i} = \frac{R_{0}}{1+n_{e}/N_{c}}
\end{equation}
with the low-density limit $R_{0}$ and the critical density $N_{c}$ where we adopted the values from Table~5 in \citet{Algol_LETGS}. Theoretical modeling of the $f/i$~ratios still suffers from the incompleteness of current atomic databases, as contributions from satellite lines and effects from dielectronic recombination have to be taken into account.

The uncertainty in the density derived from \ion{O}{vii}
is large, as a result of the low effective area at long wavelengths.
While the \ion{Ne}{ix} triplet around 13.5\AA\ is well exposed, severe
contamination complicates the line flux measurements, especially of the
\ion{Ne}{ix}~i line at 13.54~\AA\ \citep{Capella_Ne}, see also Sect.~\ref{grating}. The density value derived for EQ~Peg A is higher than
the one derived from \ion{O}{vii}. Since \ion{Ne}{ix} traces
higher coronal temperatures, this may indicate a higher pressure
in the hotter plasma regions, however, the uncertainties do
not include uncertainties from the deblending procedure.
The $f/i$-ratio derived for the
B component is -- nominally -- in the low-density limit region, however,
the error is so large that the same density derived for the A component is also
consistent with the data. The Mg and Si triplets are much stronger affected by satellite lines. Additionally, the \ion{Mg}{xi} triplet is blended with higher-order lines from the Lyman series of
\ion{Ne}{x} as discussed by \citet{HETG_density}.
The flux in the magnesium triplet lines
is too low in both components of the EQ~Peg binary for a
deblending procedure to yield any useful refinements.
Since neon turns out to be far more abundant than magnesium (see Sect.~\ref{DEM_sect}), the blending contributions
from neon amounts to a significant fraction of the line fluxes, and we consider the high
densities derived from magnesium not reliable.  Consistent with that interpretation,
the \ion{Si}{xiii} triplet lines are consistent with being
in the low-density limit, which is not surprising since
the density-sensitive range for this ion is higher than the densities normally
encountered in stellar coronae. However, the Si triplet is also problematic because of a steep increase of the effective area in both the HEG and MEG that occurs exactly at the wavelength of the forbidden line.

\begin{table}
\begin{center}
\caption{\label{densities}$f/i$ ratios (in photon units) and coronal densities for EQ~Peg~A and B from total \emph{Chandra} MEG spectra.}
\begin{tabular}{rrrrrr}
\hline\hline
&&\multicolumn{2}{c}{EQ Peg A}	& \multicolumn{2}{c}{EQ Peg B}\\
ion& log $T$	& $f/i$		& log $n_e$	& $f/i$ 	& log $n_e$\\
\hline
\ion{O}{vii} 	&6.3	&1.69$\pm$0.60	&10.62$^{+0.29}_{-0.27}$&2.81$\pm$0.87	&$< 10.98$\\
\ion{Ne}{ix} 	&6.6	&2.38$\pm$0.32	&11.40$^{+0.18}_{-0.22}$&3.53$\pm$1.33	&$< 11.51$\\
\ion{Mg}{xi} 	&6.8	&1.01$\pm$0.35	&12.99$^{+0.27}_{-0.23}$&1.68$\pm$1.04	&$< 13.28$\\
\ion{Si}{xiii} 	&6.95	&3.28$\pm$0.62	&$< 11.17$		&5.08$\pm$3.64	&$< 13.52$\\
\hline
\end{tabular}
\end{center}
\end{table}

\begin{figure}
\resizebox{\hsize}{!}{\includegraphics{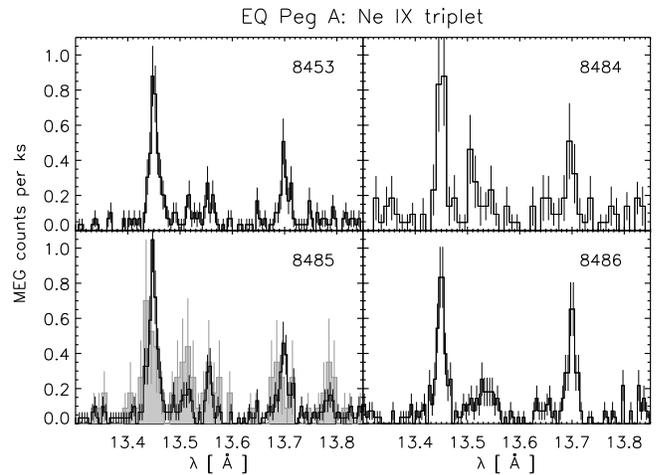}}
\caption[]{\label{Ne}The \ion{Ne}{ix} triplet in the MEG spectra of EQ~Peg~A during the four observations. On observation No. 8485 the flare spectrum only is overlaid in gray.}
\end{figure}

The flux in the \ion{Ne}{ix} triplet of EQ~Peg A is large enough to
investigate not only the density in the combined spectrum, but also
in the individual datasets. In Fig.~\ref{Ne} we show the
corresponding spectral region for each of our four data sets.
While differences in the strength of the intercombination line at
13.55~\AA\ are apparent in the individual datasets, no changes
in density can be established at a statistically significant level.
In addition to the complete dataset 8485, we show also the
spectrum extracted for the large flare (shaded in gray in the 
bottom left panel of Fig.~\ref{Ne}). 
Dividing this observation in flare and non-flaring states gives
$f/i$-ratios of 1.13$\pm$0.59 and 2.24$\pm$0.61, or density
values log~$n_e$ of 12.07$^{+0.43}_{-0.31}$ and
11.49$^{+0.32}_{-0.43}$, respectively.
For the combined quiescent spectrum, we find $f/i = 2.77\pm0.42$, corresponding to log~$n_e = 11.13^{+0.29}_{-0.09}$, which is lower but still consistent with the total spectrum (see Table~\ref{densities}) and clearly lower than during the large flare.
This indicates
activity-related density variations as observed by
\citet{Maggio_Ness_AD_Leo} on the active M~dwarf \object{AD Leo}.
Unambiguous measurements of
coronal density variations have so far only been observed during
very large flares on the active M dwarfs Proxima Centauri and
CN~Leo \citep{Guedel_Proxima_Cen_1, Coronae_Proceedings}. The
two density values derived here for flare and quiescence are
consistent with each other at the 2$\sigma$ level, however, the
density values observed during flares on active stars in
general -- although consistent with the quiescent state -- tend to result in higher densities 
\citep[see e.\,g.][]{Mitra_Kraev_densities}.

We also note the different levels of contamination of the neon triplet by \ion{Fe}{xvii} (13.8~\AA) and \ion{Fe}{xix} (13.51~\AA) in the different datasets. During the large flare, this can be attributed on the one hand to the higher coronal temperatures increasing the amount of \ion{Fe}{xix} compared to other ionization stages of iron, and on the other hand possibly also to an increased Fe/Ne abundance ratio during the flare, as discussed below.

\subsection{Differential emission measure}
\label{DEM_sect}
In order to derive absolute abundances we
investigated the differential emission measure (DEM)
of the two stars from the line fluxes of the strongest density-insensitive
lines listed in Table~\ref{counts}. For EQ~Peg A, the HEG spectrum can
also be used in addition to the MEG spectrum, while for
EQ~Peg B only the MEG spectrum provides data with a reasonable
signal-to-noise ratio. In a two-step procedure we first
reconstructed the DEM utilizing an abundance-independent
approach similar to the one proposed by \citet{Algol_EMD}. 
In a second step we then determined the abundances from the ratios of
measured line fluxes to those predicted by the DEM.

We used a volume differential emission measure defined as 
\begin{equation}
DEM(T) = n_e^2 \, \frac{dV}{dT}
\end{equation}
in units cm$^{-3}$ K$^{-1}$. The DEM is related to the theoretical photon flux $f_{ji}$ (in photons~cm$^{-2}$ s$^{-1}$) of a spectral line originating from an atomic transition from the upper level $j$ to the lower level $i$ in the ion $X^{Z+}$ by 
\begin{equation}
\label{flux}
f_{ji} = \frac{A_X}{4 \pi d^2} \int G(T) \, DEM(T) \, dT
\end{equation}
where $A_X$ is the abundance of element $X$ relative to solar photospheric values and the line contribution function $G(T)$ with
\begin{equation}
G(T) = A_{ji} \, \frac{n_j(X^{Z+})}{n(X^{Z+})} \frac{n(X^{Z+})}{n(X)} \frac{n(X)}{n(H)}_{\odot} \frac{n(H)}{n_e} \frac{1}{n_e}
\end{equation}
$A_{ji}$ being the Einstein coefficient for spontaneous emission of the transition, $n_j(X^{Z+})/n(X^{Z+})$ the level population of the upper level, $n(X^{Z+})/n(X)$ the ionization balance, $n(X)/n(H)_{\odot}$ the solar photospheric abundance, and $n(H)/n_e$ the proton-to-electron ratio, which is $\approx 0.83$ for coronal plasmas where hydrogen and helium are usually fully ionized. 
While the DEM describes the properties of the emitting plasma, the line contribution function summarizes constants and the underlying atomic physics. We used CHIANTI 5.2 \citep{Chianti7} and the ionization balance of \citet{Mazzotta_etal} to calculate $G(T)$ for each line.

Our abundance-independent reconstruction of the DEM is based
on line ratios, and we used the ratios of the H-like Ly$\alpha$ and
the He-like  resonance lines of oxygen, neon, magnesium, and
silicon, plus several ratios of adjacent ionization stages of
iron from \ion{Fe}{xvii} to \ion{Fe}{xxiv}. Since these line ratios
involve only lines of the same element, the elemental abundances
cancel out, and each ratio poses a constraint on the shape of
the DEM. The choice of these
lines yields a temperature coverage from $\approx 2$ to 20~MK,
however, the low effective area at longer wavelengths with low-temperature
lines like \ion{O}{vii} and \ion{O}{viii} results in relatively large uncertainties
at temperatures below 5~MK. For EQ~Peg A, a total of 13 ratios
have been used, while some of the ratios consisting of higher
ionization stages of iron could not be used for the B component
because of low signal, leaving only 9 usable ratios. 
In order to fix the absolute level of the DEM in addition to constraining the shape by the ratios, but still independently of a reference element like iron, 
we force the DEM and the theoretical continuum emissivity as implemented in CHIANTI to reproduce the observed continuum flux. For both components of EQ~Peg, we used the wavelength region around 7.5~\AA, where the effective area of the MEG
reaches its maximum, and the spectrum is, according to the CHIANTI database, essentially line-free, exept  for a few weak aluminum and magnesium lines. We assume that any remaining emission at these wavelengths originates from the continuum and neglect the contribution of any further weak transitions.
As a result of merging the four datasets, the determined
DEMs constitute an average over these observations with the
different states of activity and quiescent and flaring periods
of the two stars.

We used two slightly different approaches to model the DEM,
that we parameterize by polynomials of different orders.
In our first method (method~1) we fitted the polynomial
parameters as $\log DEM$, and
with our second method (method~2) we model the linear DEM,
both as a function of $\log T$. While for method~1 no boundary
conditions are needed, for method 2 we forced the value of DEM to be 
zero at two variable temperatures, and positive between these
temperatures. For a more detailed description we refer to
\citet{DEM}. 

\begin{figure}
\resizebox{\hsize}{!}{\includegraphics{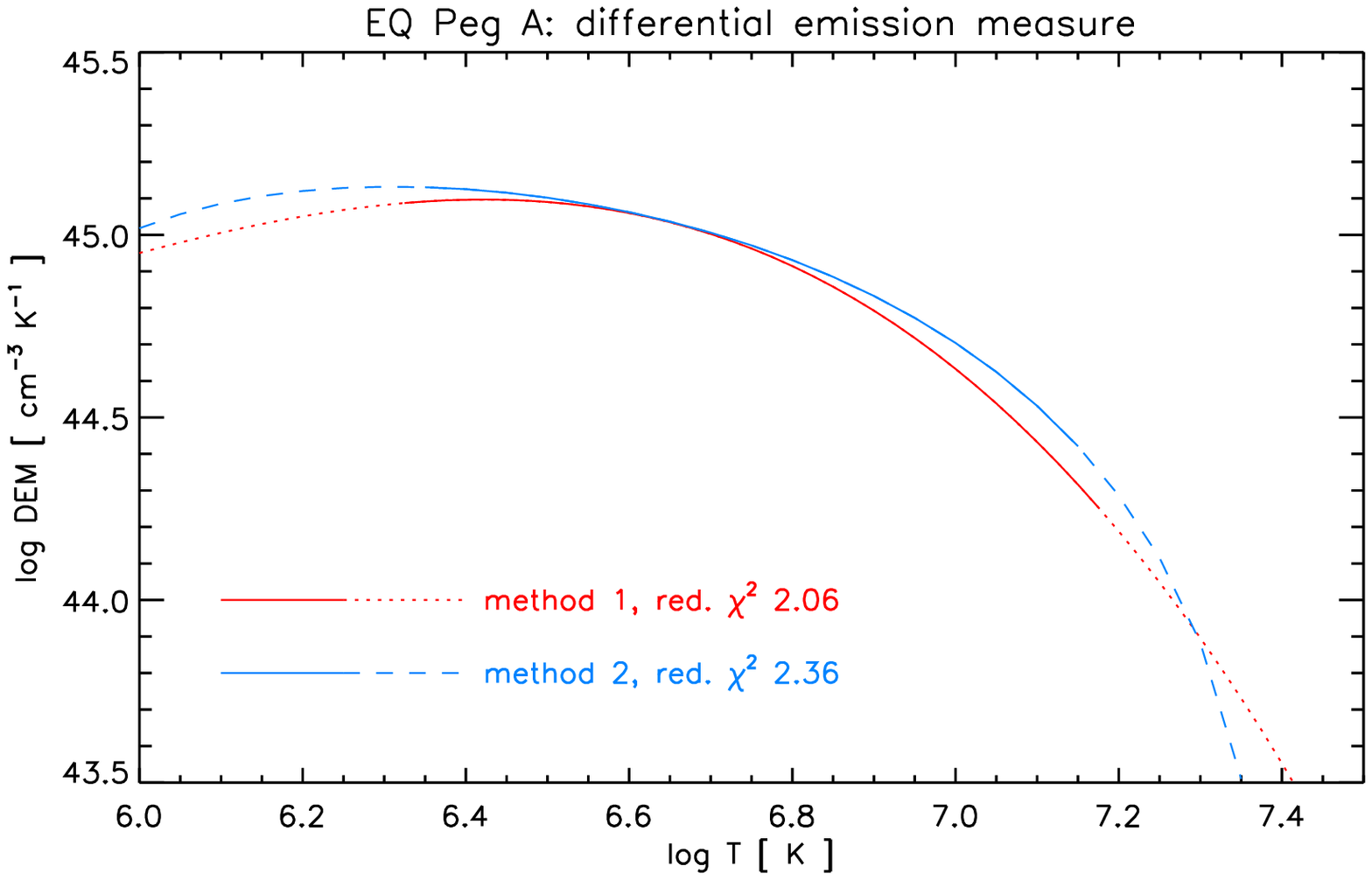}}
\resizebox{\hsize}{!}{\includegraphics{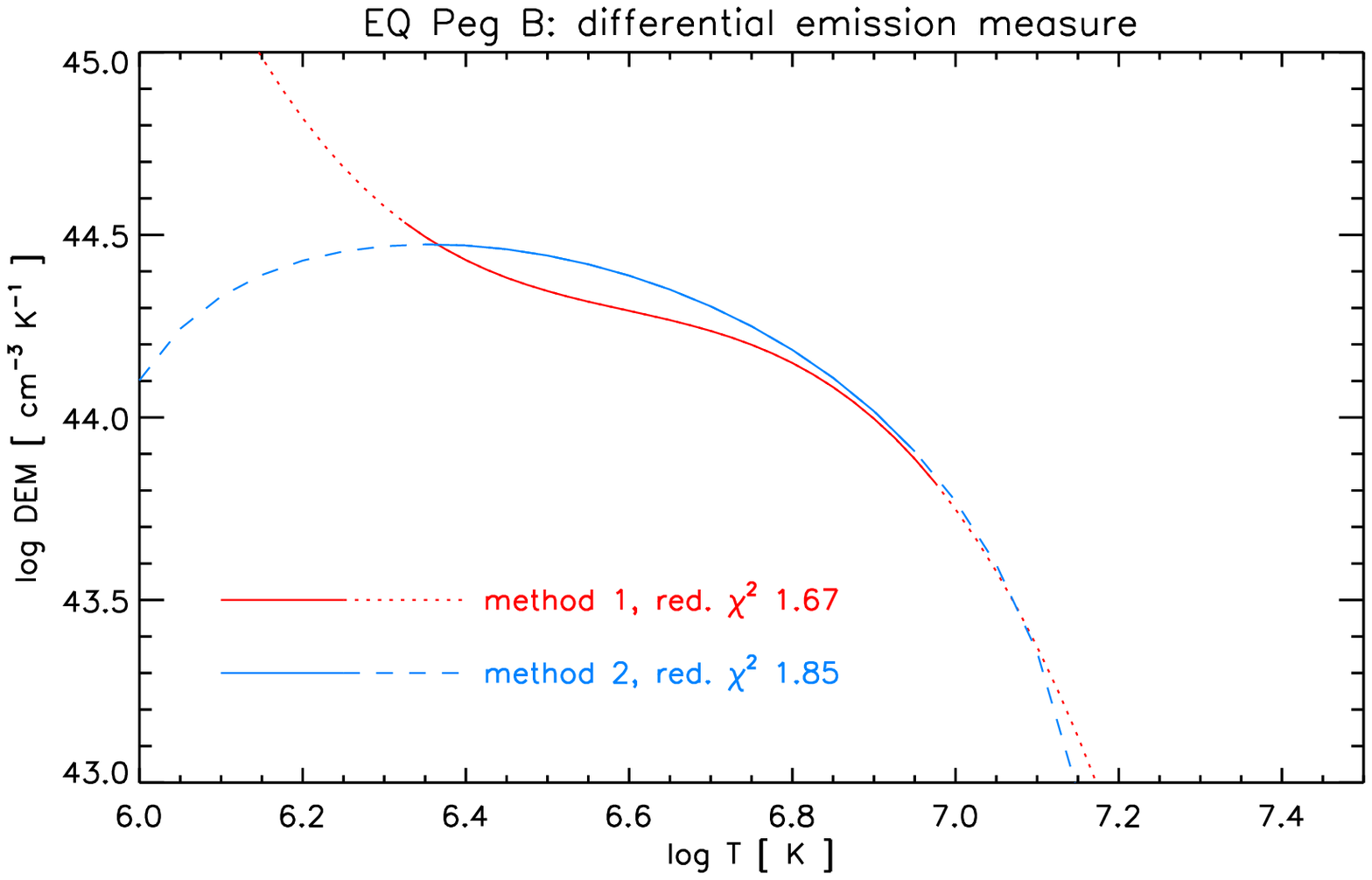}}
\caption[]{\label{DEM}Differential emission measures of EQ~Peg A (top) and B (bottom) for the total observation time (i.\,e. flare and quiescence).}
\end{figure}

For EQ~Peg A, both methods give acceptable best-fits to the
selected line ratios assuming
3$^{\rm rd}$ and 5$^{\rm th}$ order polynomials, yielding
values of reduced $\chi^2$ of 2.06 and 2.36 for methods~1 and 2,
respectively. These DEMs are shown in the top panel of 
Fig.~\ref{DEM}. The polynomials are an analytical representation of the shape of the DEM, which is only valid for certain temperature range which is at first restricted by the temperature range covered by the emissivities of the lines implemented in the fit. In Fig.~\ref{DEM} we additionally plot the polynomials with solid lines in the temperature range we consider as well-constrained from the available lines, and with dashed or dotted lines outside this range.
Since the high-temperature regime is better
represented by the available ratios than the low-temperature
range (see above), the two methods yield best agreement
at high temperatures. However, the peaks in the DEM
disagree from the two methods, and the poor constraints from
temperatures $\log T$ below 6.5 lines make it difficult to quantify the exact amount
of cool material; this may even falsely hint at a DEM peak at a lower temperature.

For EQ~Peg B, this becomes even more apparent, see the bottom
panel of Fig.~\ref{DEM}. The best-fit results are obtained with
4$^{\rm th}$ order polynomials with both methods, with reduced
$\chi^2$ values of 1.67 and 1.85; however, the two approaches
give consistent results only in the range of
$ 6.8 < \log T < 7.2$. The DEM reconstructed from method~2
shows an unphysical increase towards lower temperatures due to
the fact that no ratios are available to adequately constrain
the low-temperature slope.

When introducing higher-order polynomials (orders $> 5$), the
DEMs of both stars start to develop two-peaked structures with
maxima at $\log T \approx 6.2$--6.4 and 6.8--7.2. This would
correspond to a low- and a high temperature component, where the
first one causes the strong oxygen line fluxes, and the hotter
component is related to the typical active M dwarf corona and the
observed flares. However, the DEMs constructed with higher-order
polynomials are often unstable, and the location and strength of
certain structures often depends only on a single line flux
measurement. Thus we cannot be certain about the reality of
these structures, and we prefer the ``simpler'' DEMs with fewer
free parameters.

Next, we used the DEMs from these two approaches for the two
stars to predict line fluxes. Since the DEMs are determined
independently of the abundances and are normalized to
the continuum, any discrepancies between predictions and
measurements of line fluxes must be due to different abundances
than those assumed in predicting the line fluxes from a given
DEM \citep[here][]{Asplund}.

In Table~\ref{DEM_abund} we list our results for the abundances with statistical
uncertainties derived from the measurement uncertainties of the
line fluxes; note that the quoted errors only reflect the statistical precision
of the line flux measurements, but not the accuracy of the DEM
nor the uncertainties from the continuum flux nor any errors arising from the atomic physics parameters used. 
The errors in the continuum 
amount to 2.5\% and 8.2\% for EQ~Peg A and B, respectively, and
affect only the absolute abundance level. While we experience great
uncertainties in determining the exact shape of the DEM, the
determination of abundances seems to be only relatively little
affected. Polynomials of different orders always give abundances consistent within the errors and the abundances derived with method~1 are always consistent with those obtained from method~2. 
Similarly, \citet{ness_jordan} performed a detailed emission measure analysis of $\epsilon$~Eridani, including ionization stages from many different ions. Close to the peak, their emission measure distribution agrees with that by \citet{coronal_photospheric}, but is more accurate at low temperatures. In spite of the differences, both find similar abundances.
In general, the derived abundances turn out to be quite insensitive to changes in the shape of the DEM, polynomials of different orders always give abundance value consistent within the errors. Additionally, the abundances derived with method~1 are always consistent with those obtained from method~2. \citet{DEM} assess that the uncertainties on the abundances introduced by uncertainties on the shape of the DEM do not exceed 5\%. 


\begin{table}
\begin{center}
\caption{\label{DEM_abund}Absolute coronal abundances \citep[relative to][]{Asplund} of EQ~Peg A and B for the total observation time, determined with the DEM methods.}
\begin{tabular}{rrrrr}
\hline\hline
	&\multicolumn{2}{c}{EQ Peg A}	& \multicolumn{2}{c}{EQ Peg B}\\
element	&method 1	& method 2	& method 1	&method 2\\
\hline
N       &0.42$\pm$0.10  &0.39$\pm$0.09  &0.16$\pm$0.15  & 0.20 $\pm$0.18 \\  
O       &0.42$\pm$0.04  &0.40$\pm$0.04  &0.15$\pm$0.09  & 0.26 $\pm$0.08 \\ 
Ne      &1.14$\pm$0.07  &1.07$\pm$0.07  &0.93$\pm$0.07  & 0.83 $\pm$0.07 \\ 
Mg      &0.22$\pm$0.01  &0.20$\pm$0.02  &0.28$\pm$0.04  & 0.27 $\pm$0.04 \\ 
Si      &0.51$\pm$0.02  &0.46$\pm$0.02  &0.54$\pm$0.07  & 0.52 $\pm$0.07 \\ 
S       &0.64$\pm$0.16  &0.58$\pm$0.21  &---            & ---            \\
Fe      &0.19$\pm$0.02  &0.18$\pm$0.01  &0.20$\pm$0.02  & 0.17 $\pm$0.02 \\ 
\hline                                                   
\end{tabular}                                            
\end{center}                                             
\end{table} 
                                             
In Table~\ref{DEM_abund} we show that the absolute abundances relative to\citet{Asplund} of
EQ~Peg A and B are roughly consistent for all elements, yielding
a low metallicity compared to solar. This is typical for active
M dwarfs \citep[compare e.\,g. with the values derived by][from
global fitting approaches to CCD X-ray spectra]{Robrade_M_dwarfs}.
Nitrogen, oxygen, and neon may be somewhat
more abundant in the A component, however, not at a statistically
significant level; also note that the uncertainties for N and O
are quite large.  While the high-FIP element neon is about
solar, the low-FIP element iron shows the lowest abundance value,
and the other elements also follow the typical inverse FIP effect. 


The computed abundances and DEMs can now be used to model the spectra of EQ~Peg A and B. We list the calculated theoretical line fluxes together with the measured fluxes in Table~\ref{counts}.

\subsection{Emission measure-independent abundances}
\begin{figure}
\resizebox{\hsize}{!}{\includegraphics{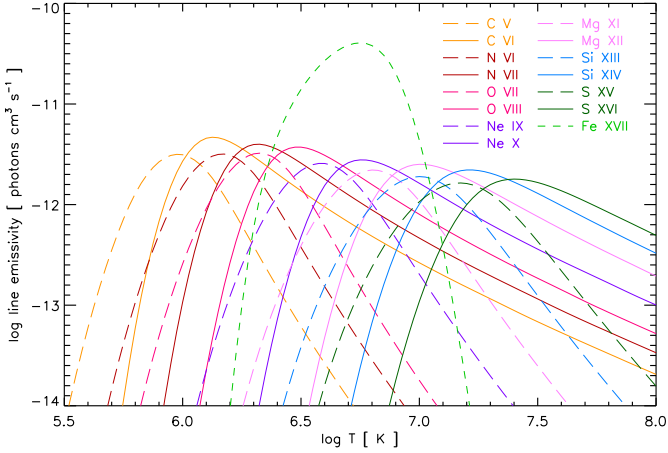}}
\caption[]{\label{gofts}Line emissivity curves for different H-like and He-like ions, and \ion{Fe}{xvii} at 15.01~\AA\ from CHIANTI 5.2 \citep{Chianti7} with the ionization balance from \citet{Mazzotta_etal}.}
\end{figure}

Given the ill-posed nature of the DEM determinations and the
resulting uncertainties in the DEM distributions, we also investigated
emission measure-independent abundance determinations.
The strongest lines in stellar coronal X-ray spectra (and thus 
easy to measure even in spectra with low signal) originate from 
the H-like and He-like lines of carbon, nitrogen, oxygen, neon, 
magnesium, and silicon, as well as Ne-like \ion{Fe}{xvii}. In
Fig.~\ref{gofts} we show the line emissivities for these ions as
a function of temperature. While these lines are formed over a 
broad temperature range, their emissivity peaks are
relatively narrow. Those lines for which the line
contribution functions have similar peak formation temperatures
yield a similar temperature dependence. Ratios of their
line fluxes depend only weakly on temperature,
but strongly on the abundance ratio of the respective elements. 
Already \citet{Acton_neon_to_oxygen} proposed to determine the
solar coronal Ne/O abundance ratio from the ratio of the
measured fluxes of the \ion{Ne}{ix} resonance and
\ion{O}{viii} Ly $\alpha$ lines. This method yields a comparably high
precision (because of smaller measurement uncertainties),
and also a higher accuracy (because no systematic
uncertainties from the DEM are introduced).
Recent examples of such approaches were presented by
\citet{CNO} and \citet{Drake_Testa}.

While for a given element, the peak formation temperature of
the H-like ion always exceeds that of the corresponding He-like ion,
the peak formation temperatures of the H-like and He-like ions
shift to higher temperature with increasing atomic mass.
Therefore, pairs of H-like and He-like ions of elements with
different atomic mass yield similar peak formation temperatures,
and their contribution functions have similar shapes. Further examples
for such pairs are the H-like lines and He-like lines of carbon and nitrogen,
nitrogen and oxygen, neon and magnesium, and magnesium and
silicon respectively \citep{Parkinson_abundance_ratios}. However, these 
ratios are not perfectly
temperature-independent, and \citet{Drake_Testa} have refined
Acton's method by computing a linear combination of
\ion{Ne}{x} Ly~$\alpha$ and the \ion{Ne}{ix} r line in order to
construct a new contribution function of Ne lines that is more
similar in shape to the \ion{O}{viii} Ly~$\alpha$ line. 

This suggests to ask in general, which linear combinations of
strong lines without significant blends (i.\,e. hydrogen- 
and helium-like lines and the strongest lines of \ion{Fe}{xvii} 
as mentioned above) of two certain elements yield the smallest 
temperature residuals in their normalized contribution functions.
We thus write for the abundance ratio of two elements with respective abundances
$A_1$ and $A_2$
\begin{equation}
\label{lincomb}
\frac{\mathrm{A}_1}{\mathrm{A}_2} = \frac{f(\mathrm{A_1^{Z-1}}) + C_1 \cdot f(\mathrm{A_1^{Z}})}{ C_2 \cdot f(\mathrm{A_2^{Z-1}}) + C_3 \cdot f(\mathrm{A_2^{Z}})},
\end{equation}
with $f(\mathrm{A_1^{Z-1}})$ and $f(\mathrm{A_2^{Z-1}})$ denoting the
measured line fluxes $f$ of the He-like resonance lines and
$f(\mathrm{A_1^{Z}})$, and $f(\mathrm{A_2^{Z}})$ the
corresponding H-like Ly$\alpha$ lines
of two elements A$_1$ and A$_2$, respectively, and 
\begin{equation}
\label{lincomb_Fe}
\frac{\mathrm{A}}{\mathrm{A_{Fe}}} = \frac{ C_4 \cdot(f(\mathrm{A^{Z-1}}) + C_5 \cdot f(\mathrm{A^{Z}}))}{ \sum f(\mathrm{\ion{Fe}{xvii}})}
\end{equation}
where $\sum f(\ion{Fe}{xvii})$ corresponds to $f(\ion{Fe}{xvii}~15.01~\mathrm{\AA}) + f(\ion{Fe}{xvii}~16.78~\mathrm{\AA}) + f(\ion{Fe}{xvii}~17.05~\mathrm{\AA}) +  f(\ion{Fe}{xvii}~17.09~\mathrm{\AA})$, to determine relative 
abundance ratios. 

To obtain the coefficients $C_1$, $C_2$, $C_3$, $C_4$, and $C_5$, we performed a minimization of the temperature residuals of the corresponding linear combination of the theoretical emissivities $\epsilon$ for the involved lines over a given temperature range:
\begin{equation}
\label{minimum}
\chi^2 = \sum_i \Bigl(\epsilon_i(\mathrm{A_1^{Z-1}}, T_i) + C_1 \cdot \epsilon_i(\mathrm{A_1^{Z}}, T_i) - C_2 \cdot \epsilon_i(\mathrm{A_2^{Z-1}}, T_i) - C_3 \cdot \epsilon_i(\mathrm{A_2^{Z}}, T_i)\Bigr)^2
\end{equation}
for the coefficients in Eqn~\ref{lincomb}, and 
\begin{equation}
\label{minimum_Fe}
\chi^2 = \sum_i \Bigl(\epsilon_i(\mathrm{A^{Z-1}}, T_i) + C_5 \cdot \epsilon_i(\mathrm{A^{Z}}, T_i) - \frac{1}{C_4} \cdot {\textstyle\sum} \epsilon_i(\mathrm{\ion{Fe}{xvii}}, T_i)\Bigr)^2
\end{equation}
for the coefficients in Eqn~\ref{lincomb_Fe}.

In Table~\ref{coeff} we list these coefficients for the temperature range where the line emissivities of all involved lines exceed 1\% of their peak emissivity. In the given form, the coefficients convert the measured line fluxes in photon flux units to absolute abundance ratios, i.\,e. independent from any set of solar photospheric abundances; we used line emissivities from CHIANTI 5.2 \citep{Chianti7}. 

\begin{table}
\begin{center}
\caption{\label{coeff}Coefficients for Eqns~\ref{lincomb} and \ref{lincomb_Fe} to obtain temperature-independent abundance ratios.} 
\begin{tabular}{lccccc}
\hline\hline
ratio  & $C_1$ & $C_2$ & $C_3$ & $C_4$ & $C_5$\\
\hline
 N / C      & +0.13 & $-0.07$ & +0.73& --- &--- \\
 O / N 	    & +0.30 & $+0.01$ & +0.93& --- &---\\
 Ne / O     & +0.02 & $-0.17$ & +0.69& --- &---\\
 Mg / Ne    & +0.18 & $-0.08$ & +0.87& --- &---\\ 
 Si / Mg    & +0.32 & $+0.05$ & +0.86& --- &---\\  
 S / Si     & +0.42 & $+0.15$ & +0.85& --- &---\\
 Ne / Fe    & ---  & ---     & --- & +34.71 & +0.46\\ 
 Mg / Fe    & ---  & ---     & --- & +67.73 & $-0.30$\\
\hline
\end{tabular}
\end{center}
\end{table}

\begin{figure}
\resizebox{\hsize}{!}{\includegraphics{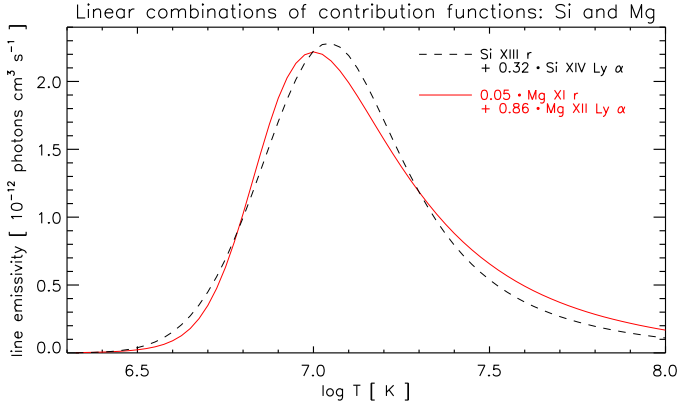}}
\caption[]{\label{Si_Mg}Best-fit linear combination of the contribution functions of the H-like Ly~$\alpha$ and He-like resonance lines of silicon and magnesium.}
\end{figure}

There are still residuals in temperature for these linear
combinations, and their amplitudes differ for each ratio. We
found the lowest residuals for the Si/Mg ratio and the
largest ones for the Ne/O ratio. In Fig.~\ref{Si_Mg} we
illustrate the best-fit linear combination of emissivities
that yield the Si/Mg abundance ratio, and for the Ne/O ratio
we refer to Fig.~3 of \citet{alpha_Cen_Ne/O}, who had used
the same coefficients as listed here to calculate the Ne/O abundance ratio
of $\alpha$~Centauri. The first six ratios involve only H-like
Ly~$\alpha$ and He-like resonance lines. However, the S/Si
ratio should only be used when the sulfur lines can be measured
with reasonable signal-to-noise in \emph{Chandra} HETGS spectra.
\ion{Fe}{xvii} can
be matched with linear combinations of hydrogen- and helium-like
lines of neon or magnesium. The residuals are much smaller for
Mg, however, the Mg lines are often weak, introducing large
measurement uncertainties. Especially the weak \ion{Mg}{xi}
line carries a large weight, and for the determination of Fe abundances we prefer
the Ne/Fe ratio despite its larger residuals.

The coefficients of the linear combinations as listed in
Table~\ref{coeff} can be adjusted in some cases where certain
line fluxes are not available (e.\,g. the \ion{C}{v} lines can
only be measured with the \emph{Chandra} LETGS), or are extremely
weak (e.\,g. \ion{Mg}{xi} or \ion{N}{vi}, depending on temperature). 
The line fluxes $f(\mathrm{A_1^{Z-1}})$ and $f(\mathrm{A_2^{Z}})$
with Z(A$_1)>$Z(A$_2$) provide the major constituents of the linear 
combinations for the ratios not involving Fe.

For example, the \ion{N}{vi} resonance line is often weak in stars with hot coronae and is difficult to
measure in \emph{Chandra} MEG spectra, however, this line contributes much less to the nitrogen linear
combination in the O/N ratio than the \ion{N}{vii} Ly~$\alpha$ line (i.\,e. $C_2$ is much smaller than $C_3$). We readjusted $C_1$, 
and $C_3$ assuming the fixed value $C_2=0$ and found only
marginal changes for the values of the other coefficients listed in Table~\ref{coeff}.
Although the value of $C_1$ is also quite low for the
\ion{Ne}{x} Ly~$\alpha$ line in the Ne/O ratio, the
situation is somewhat different. If we readjust the other
coefficient after setting $C_1=0$, the residuals are much
larger, especially at high temperatures where the emission
measure distributions of active stars usually have their
maximum.

Finally, the coefficients can be re-computed
for the sum of the He-like resonance, intercombination, and
forbidden lines for cases where these lines can not be
resolved (e.\,g. \ion{Si}{xiii} with the \emph{Chandra}
LETGS).

We applied Eqns~\ref{lincomb} and \ref{lincomb_Fe}, with the coefficients from
Table~\ref{coeff} multiplied by the solar photospheric abundances of \citet{Asplund}, to the line fluxes measured from the spectra of  EQ~Peg A (including and excluding periods of
flare activity) and of EQ~Peg B and list the resulting abundance ratios in Table~\ref{ratios}.

Within the uncertainties, the derived abundance ratios are
consistent for all three datasets. Only the Ne/Fe ratio is
conspicuously lower
in the quiescent spectrum of EQ~Peg~A compared to the total
spectrum, and again lower for the more inactive EQ~Peg~B.
Flare- or activity-related changes in the abundance ratios
cannot be determined from count statistics; however, if real, the
differences may be caused by a decreased iron abundance level
or a higher neon abundance on the more active A component,
especially during flares.

\begin{table}
\begin{center}
\caption{\label{ratios}Abundance ratios \citep[relative to][]{Asplund} applying temperature structure-independent linear combinations of coronal emission lines. Values given in brackets for EQ~Peg~A correspond to the quiescent state.}
\begin{tabular}{cccc}
\hline\hline
ratio	& \multicolumn{2}{c}{EQ Peg A} & EQ Peg B\\
\hline
O / N 	& 0.83$\pm$0.20 & (1.00$\pm$0.29) & 0.81$\pm$0.72\\
Ne / O  & 2.48$\pm$0.31 & (2.02$\pm$0.16) & 2.18$\pm$0.41\\
Mg / Ne & 0.20$\pm$0.02 & (0.19$\pm$0.02) & 0.32$\pm$0.07\\
Si / Mg & 2.52$\pm$0.21 & (2.56$\pm$0.25) & 2.05$\pm$0.44\\
S / Si	& 1.26$\pm$0.20 &---		& ---\\
Ne / Fe	& 7.30$\pm$ 0.64& (5.99$\pm$ 0.34)&5.45 $\pm$ 0.78 \\      
Mg / Fe	& 1.27$\pm$ 0.17& (1.09$\pm$ 0.19)&1.63$\pm$  0.43\\  
\hline
\end{tabular}
\end{center}
\end{table}

In Table~\ref{DEM_ratios} we list the abundance ratios formed from the absolute values listed in Table~\ref{DEM_abund} for comparison. They are consistent within the 
errors with the values in Table~\ref{ratios}, but the ratios obtained from the emission 
measure-independent method are more robust and have lower uncertainties. 

\begin{table}
\begin{center}
\caption{\label{DEM_ratios}Abundance ratios relative to \citet{Asplund} of EQ~Peg A and B, determined from the absolute abundances listed in Table~\ref{DEM_abund}.}
\begin{tabular}{ccccc}
\hline\hline
	&\multicolumn{2}{c}{EQ Peg A}	& \multicolumn{2}{c}{EQ Peg B}\\
ratio	&method 1	& method 2	& method 1	&method 2\\
\hline
O / N 	&1.00$\pm$0.25  &1.03$\pm$0.26  &0.94$\pm$0.94  & 1.30 $\pm$1.24 \\   
Ne / O  &2.71$\pm$0.31  &2.68$\pm$0.32  &6.20$\pm$3.75  & 3.19 $\pm$0.94 \\  
Mg / Ne &0.19$\pm$0.01  &0.19$\pm$0.02  &0.30$\pm$0.05  & 0.33 $\pm$0.06 \\  
Si / Mg &2.32$\pm$0.14  &2.30$\pm$0.25  &1.93$\pm$0.37  & 1.93 $\pm$0.39 \\      
S / Si  &1.25$\pm$0.32  &1.26$\pm$0.46  &---    	& ---            \\      
Ne / Fe &6.00$\pm$0.73  &5.94$\pm$0.51  &4.65$\pm$0.58  & 4.88 $\pm$0.71 \\     
Mg / Fe &1.15$\pm$0.13  &1.11$\pm$0.13  &1.40$\pm$0.24  & 1.59 $\pm$0.30 \\    
\hline                                                   
\end{tabular}                                            
\end{center}                                             
\end{table}

\section{EQ Peg in the context of active M dwarfs}
In order to put EQ~Peg into the context of other similar late-type stars, we
extracted five additional \emph{Chandra} HETGS spectra of active M
dwarfs for comparison with our results; our sample thus
contains \object{YY Gem}, \object{AU Mic}, \object{EV Lac}, \object{AD Leo}, and
\object{Proxima Cen} in addition to EQ~Peg~A and B. Most of these archival
observations have been included in many studies
\citep[e.\,g.][]{opacity, sizes, Testa_optical_depth}, the
X-ray lightcurves and spectra of AU~Mic, EV~Lac, and
Proxima~Cen are shown by \citet{HETG_density}.
While the two stars in the binary EQ~Peg are co-eval, we are now
dealing with a range of different ages, rotation periods, and metallicity. 
In Table~\ref{M_dwarfs} we summarize for each star their spectral
type, distance (from the SIMBAD database), age estimates and rotation periods from the literature 
together with observation ID, exposure time,
and X-ray luminosity obtained from integration of the dispersed
photons over the range 2--25\,\AA. The bolometric luminosity for $L_{\mathrm{X}}/L_{\mathrm{bol}}$ has been calculated according to \citet{Kenyon_Hartmann}. While the X-ray luminosity decreases along the sequence of spectral types, $L_{\mathrm{X}}/L_{\mathrm{bol}}$ is in the saturation regime of $\approx -3.3$ for all stars except Proxima Cen.

\begin{table*}
\begin{center}
\caption{\label{M_dwarfs}Properties of active M dwarfs observed with \emph{Chandra} HETGS. X-ray luminosities are computed from the total background-subtracted counts contained in the MEG spectra from 2--25~\AA. Values given in brackets correspond to the quiescent state.}
\begin{tabular}{lccccccccccc}
\hline\hline
star		& spectral& distance&age&P$_\mathrm{rot}$& ObsID	&\multicolumn{2}{c}{exposure time}&\multicolumn{2}{c}{log $L_{\mathrm{X}}$}&\multicolumn{2}{c}{log $L_{\mathrm{X}}/L_{\mathrm{bol}}$}\\
		& type	& [ pc ]& [ Myrs ]	& [ d ]		& &\multicolumn{2}{c}{[ ks ]}			&\\
\hline
YY Gem		& M\,0.5&15.27	&370$ ^1$	& 0.814$ ^5$	&8504/8509		&136.0 &(81.5)	& 29.63 &(29.43)	&$-3.17$&($-3.37$)\\
AU Mic 		& M\,1	& 9.94 	&12$ ^2$	& 4.865$ ^6$	&17 			&58.8 &(47.4)	& 29.19 &(29.13)	&$-3.48$&($-3.54$)\\
EQ Peg A	& M\,3.5& 6.25 	&950$ ^3$	& (1.066$ ^7$)	& 8453/8484/8485/8486	&98.5 &(92.8)	& 28.71 &(28.67)	&$-3.19$&($-3.23$)\\
EV Lac 		& M\,3.5& 5.05 	&45$ ^3$	& 4.376$ ^8$	& 1885			&100.0 &(87.3)	& 28.63 &(28.51)	&$-3.10$&($-3.22$)\\
AD Leo 		& M\,4.5& 4.69 	&470$ ^3$	& 2.7/14$ ^{9,10}$& 2570			&45.2	&---	& 28.54&---		&$-3.36$&---\\
EQ Peg B	& M\,4.5& 6.25 	&950$ ^3$	& (1.066$ ^7$)	&8453/8484/8485/8486		&98.5	&---	& 27.89&---	&$-3.39$&---\\
Proxima Cen 	& M\,5.5& 1.30 	&5800$ ^4$	& 83.5$ ^{10}$	&2388 			&42.4 &(35.7)	& 26.85 &(26.60)	&$-3.96$&($-4.21$)\\
\hline
\end{tabular}
\end{center}
$ ^1$ \citet{Torres_YY_gem}, $ ^2$ \citet{Zuckerman_beta_Pic_group}, $ ^3$ \citet{Parsamyan_ages}, $ ^4$ \citet{Yildiz_alpha_Cen} , $ ^5$ \citet{Kron_P_rot_YY_Gem}, $ ^6$ \citet{Torres_P_rot_AU_Mic}, $ ^7$ \citet{Norton_P_rot_EQ_Peg},  $ ^8$ \citet{Pettersen_P_rot_EV_Lac},  $ ^9$ \citet{Spiesman_P_rot_AD_Leo}, $ ^{10}$ \citet{Mahmoud_P_rot_AD_Leo}, $ ^{11}$ \citet{Benedict_P_rot_Proxima}
\end{table*}

YY Gem is an M0.5 companion of the Castor AB system. YY~Gem itself is an eclipsing binary consisting of two almost identical early M~dwarfs with an orbital period of 0.814 days at an inclination of 86.2, separated by 3.88~R$_{\sun}$. From eclipse mapping techniques based on an \emph{XMM-Newton} observation, \citet{Guedel_YY_Gem} found similar activity levels on both components, but concentrated in compact active regions at small scale heights, i.\,e. inter-binary loops as discussed for RS~CVn systems are rather unlikely. This is confirmed by \citet{Stelzer_YY_Gem}, who modelled a flare in a simultaneous \emph{XMM-Newton} and \emph{Chandra} LETGS observation. Several strong flares occur also in the two HETGS observations discussed here.
The young ($\approx 10$~Myrs) star AU~Mic, spectral type M1,
has the highest X-ray luminosity within 10~pc to the Sun and has
thus been extensively observed in X-rays and in the EUV. AU~Mic
shows a strong level of variability on all time scales and a high
flare rate. From the HETGS dataset used here,
\citet{Linsky_AU_Mic} computed an emission measure distribution
with a peak temperature of $\log T \approx 6.8$ and subsolar
abundances with a pronounced inverse FIP effect.
\citet{Magee_AU_Mic} found similar values for the abundances based
on an \emph{XMM-Newton} observation. The \emph{Chandra} HETGS
observation of the M3.5 star EV~Lac is part of a multiwavelength
campaign and has been discussed in detail by
\citet{Osten_EV_Lac_flare,Osten_EV_Lac_quiescence}. Several
flares occurred towards the end of the observation, and
\citet{Osten_EV_Lac_flare} compared line ratios with different
FIP dependencies during flares and quiescence and find a slight
increase for low FIP/high FIP ratios during flares.
\citet{Osten_EV_Lac_quiescence} derive a subsolar abundance
level and an inverse FIP effect from an emission measure
distribution of the quiescent state, with values consistent with
an \emph{XMM} observation analyzed by \citet{Robrade_M_dwarfs}. 
Both EV~Lac and the M4.5 star AD~Leo are young stars from the
galactic disk population. AD~Leo is another a well-known flare
star \citep[e.\,g.][]{Hawley_Allred}, however, no larger flares
have occurred during the \emph{Chandra} HETGS observation
discussed here. This dataset was analyzed by
\citet{Maggio_Ness_AD_Leo}, with the focus on density
diagnostics. \citet{Maggio_AD_Leo} found a mild inverse FIP
bias based on two \emph{Chandra} LETGS observations, which is
roughly consistent with the results obtained by
\citet{Robrade_M_dwarfs} and from an \emph{XMM-Newton} observation. 
\citet{vandenBesselaar_AD_Leo} analyzed the same data. 
While \citeauthor{Robrade_M_dwarfs} and \citeauthor{vandenBesselaar_AD_Leo}
determined an overall subsolar abundance level, 
\citeauthor{Maggio_AD_Leo} found values much higher than for
the solar photosphere, however with a large uncertainty in the
overall normalization. EQ~Peg~A, EV~Lac, and AD~Leo have
comparable X-ray luminosities. On the basis of a common origin
with the $\alpha$~Centauri system, Proxima Cen can be considered
older than the Sun. With a spectral type of M5.5, it is also
the latest star in our sample. Its X-ray luminosity is comparable
to the Sun, more than two orders of magnitude lower than that
of AU~Mic. Nevertheless, Proxima~Cen is also a well-known flare
star. \citet{Guedel_Proxima_Cen_1} and
\citet{Guedel_Proxima_Cen_2} discussed a large-amplitude
long-duration flare observed with \emph{XMM-Newton}.
\citet{Guedel_Proxima_Cen_2} and \citet{Nordon_flares2} derived
emission measure distributions for the different phases of the
flare and found abundance ratios relative to iron to stay
roughly constant, but with an overall tendency to an inverse
FIP effect in the absolute level. Two flares are also included
in the HETGS observation analyzed here.

In order to obtain self-consistent results, we applied the same data 
reduction and analysis techniques as
described above to the spectra of the other four sample stars.
However, even with typical flare rates and amplitudes for
active M~dwarfs known, the occurrence of individual flares is
randomly distributed, and each X-ray observation is a snapshot
that is not necessarily representative of the typical flaring
behavior of the observed star. We filtered the observation
for quiescent-only emission and include the quiescent
luminosities in brackets in the last column of
Tables~\ref{M_dwarfs}.

\begin{figure}
\resizebox{\hsize}{!}{\includegraphics{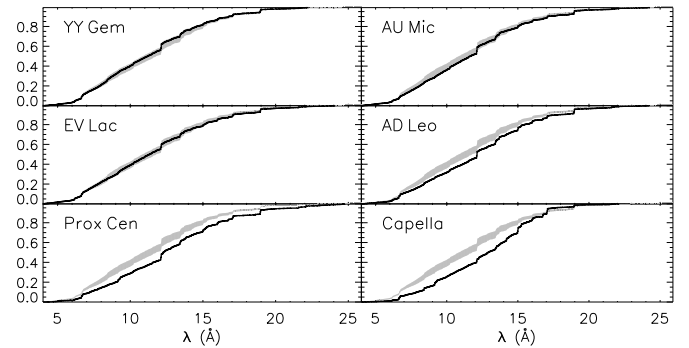}}
\caption[]{\label{accum}Cumulative distribution of counts
in the quiescent spectra of the reference star Capella and other M~dwarfs observed with the
\emph{Chandra} HETGS. The area between the curves for EQ~Peg A
and B is indicated in shades of gray for comparison.}
\end{figure}

We also computed the cumulative count spectra for the other five
spectra as that shown in Fig.~\ref{cumulative} for EQ~Peg.
In Fig.~\ref{accum} we show these spectra for the
quiescent states of the five stars as labeled in the legends, together with the well-exposed reference spectrum of Capella.
For comparison, we included the cumulative spectra of EQ~Peg~A
and B, bracketing the gray-shaded areas. In these plots, YY~Gem, AU~Mic and
EV~Lac are similar to the EQ~Peg system, with YY~Gem very similar to
EQ~Peg~A, while AU~Mic has stronger neon lines than EQ~Peg, leading 
to larger steps in the cumulative distribution. AD~Leo is more similar to
EQ~Peg B, and Proxima~Cen is even softer.
All seven M~dwarfs differ considerably from Capella.

We applied our DEM-independent abundance determination
using Eq.~\ref{lincomb} with the coefficients listed in
Table~\ref{coeff} to the total and quiescent-only spectra
of AU~Mic, EV~Lac, AD~Leo, and Proxima~Cen and list the
results in Table~\ref{M_dwarf_ratios}. These numbers have
to be compared to the relative abundances for EQ~Peg A and
B listed in Table~\ref{ratios}. 
For comparison of our results with other abundance determinations for our sample stars, we transformed the values determined by \citet{Robrade_M_dwarfs}, \citet{Guedel_YY_Gem}, \citet{Linsky_AU_Mic}, \citet{Magee_AU_Mic}, \citet{Osten_EV_Lac_quiescence}, \citet{Maggio_AD_Leo}, and \citet{Guedel_Proxima_Cen_2} to the reference set of solar photospheric abundances from \citet{Asplund} and found good agreement within the errors. Only for neon we found a slight tendency of overestimating the abundance, as the blends to the neon lines may not always have been taken properly into account.

\begin{figure}
\resizebox{\hsize}{!}{\includegraphics{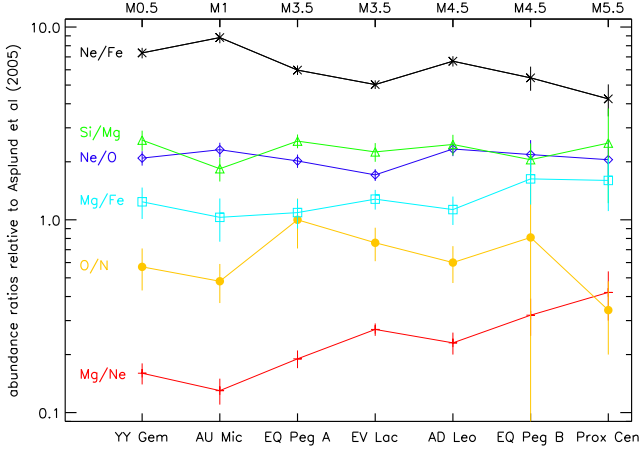}}
\caption[]{\label{abundances}Relative abundances as a function of
spectral type derived from the seven M dwarfs as listed in
Tables~\ref{ratios} and \ref{M_dwarf_ratios}.}
\end{figure}

For all seven stars, the
Ne/O ratio is enhanced by about a factor of 2, which is well-consistent with what \citet{Drake_Testa} found for their sample of active stars. Si/Mg is also
enhanced by a factor of about 2, while Mg/Ne is depleted by 
factors from 3 (EQ~Peg B and Proxima~Cen) up to 10 (AU~Mic).
The values for O/N range from 0.3--1.0 times the solar level, 
partly with large uncertainties. The Ne/Fe ratio
is clearly increased, by factors ranging from 4.5 (Proxima~Cen)
to 9 (AU~Mic), while the Mg/Fe ratio is about solar. In 
Fig.~\ref{abundances} we give a graphical representation of the
numbers listed in Tables~\ref{ratios} and \ref{M_dwarf_ratios}
from the quiescent data.

\begin{table*}
\begin{center}
\caption{\label{M_dwarf_ratios}Abundance ratios of other M dwarfs observed with the \emph{Chandra} HETGS. Values obtained for the quiescent state are given in brackets.}
\begin{tabular}{cccccccccc}
\hline\hline
ratio	&\multicolumn{2}{c}{YY Gem} & \multicolumn{2}{c}{AU Mic}	&  \multicolumn{2}{c}{EV Lac}	& AD Leo	& \multicolumn{2}{c}{Proxima Cen}\\
\hline
O / N 	& 0.62$\pm$0.12&(0.57$\pm$0.14)& 0.48$\pm$0.09 &(0.48$\pm$0.11)& 0.53$\pm$0.10 &(0.76$\pm$0.15)& 0.60$\pm$0.13 & 0.30$\pm$0.09&(0.34$\pm$0.14)\\
Ne / O  & 1.97$\pm$0.12&(2.10$\pm$0.18)& 2.36$\pm$0.19 &(2.31$\pm$0.20)& 2.07$\pm$0.50 &(1.71$\pm$0.12)& 2.33$\pm$0.19 & 1.72$\pm$0.31&(2.06$\pm$0.54)\\
Mg / Ne & 0.18$\pm$0.02&(0.16$\pm$0.02)& 0.11$\pm$0.01 &(0.12$\pm$0.02)& 0.24$\pm$0.02 &(0.22$\pm$0.02)& 0.21$\pm$0.02 & 0.31$\pm$0.06&(0.35$\pm$0.10)\\
Si / Mg & 2.31$\pm$0.18&(2.59$\pm$0.31)& 2.33$\pm$0.26 &(1.84$\pm$0.26)& 2.47$\pm$0.22 &(2.25$\pm$0.25)& 2.46$\pm$0.30 & 2.48$\pm$0.28&(2.50$\pm$1.28)\\
S / Si  & 1.27$\pm$0.18&(1.46$\pm$0.33)&	---	& ---		& 1.82$\pm$0.33 &---		&	---   &	---	& ---		\\
Ne / Fe	& 7.08$\pm$0.33&(7.34$\pm$0.46)& 8.85$\pm$0.57 &(9.03$\pm$0.59)& 5.73$\pm$0.29 &(5.17$\pm$0.26)& 6.78$\pm$0.41 & 4.44$\pm$ 0.59&(4.45$\pm$ 0.82)\\   
Mg / Fe	& 1.23$\pm$0.19&(1.24$\pm$0.24)& 0.84$\pm$0.18 &(1.03$\pm$0.26)& 1.28$\pm$0.15 &(1.28$\pm$0.15)& 1.13$\pm$0.19 & 1.44$\pm$ 0.35&(1.60$\pm$ 0.49)\\     
\hline  
\end{tabular}
\end{center}
\end{table*}

O/N and Ne/O can be considered as high~FIP/high~FIP ratios, and
within the errors, these ratios do not change with spectral type.
No trends can also be identified for the low~FIP/low~FIP ratios
Mg/Fe and Si/Mg. The low~FIP/high~FIP ratio Mg/Ne on the other hand
clearly increases towards later spectral type, while the 
high~FIP/low~FIP ratio Ne/Fe decreases. 

Mg/Ne and Ne/Fe are sensitive to a possible FIP bias, and both
are anomalously low and high, respectively, for all stars in our
sample, pointing towards an inverse FIP effect. This result holds for both
total and quiescent-only spectra. Since we only have relative
abundances, it is unclear whether high-FIP elements are
overabundant or low-FIP elements are underabundant. A low Mg/Ne
ratio may be caused by an enhancement of neon, as supported by the 
increased Ne/O ratios, but also by a low Mg abundance, which would 
match the enhanced Si/Mg ratios, and so on.

The trends observed for Ne/Fe and Mg/Ne are roughly independent of the occurrence of flares. This confirms previous findings that strength of the inverse FIP effect scales with the general activity level of a star \citep{Guedel_FIP_IFIP, Sun_in_time}, as activity (in terms of X-ray luminosity) decreases with increasing spectral type. While the behavior of the Ne/Fe and Mg/Ne ratios in our M~dwarf sample implies an absolute decrease of neon with later spectral type, the constant Ne/O ratio does not fit that picture. This indicates that not only a single element like neon causes this effect, but that the inverse FIP effect in general diminishes for later M~dwarfs.

\section{Discussion}
The behavior of M dwarf coronae shows striking changes with spectral type. The early M dwarfs up to a spectral type of $\approx$~M3--M4 observed so far are strong X-ray emitters, showing permanent flaring, high coronal temperatures and a pronounced inverse FIP effect \citep{Robrade_M_dwarfs}. X-ray luminosity, the level and peak temperature of the emission measure, and the flaring rate decrease for later M~dwarfs. While for earlier M~subclasses, it is difficult  -- if not impossible -- to distinguish between weak flaring and quiescence, M~dwarfs of spectral type M5 or later show periods of quiescence interrupted by discrete flares, and their coronal temperatures decrease \citep{Fuhrmeister_CN_Leo}. Very late M dwarfs ($>$M7) become more and more difficult to detect as quiescent X-ray sources, yet they can still produce enormous X-ray luminosities during flares.  Abundance determinations are difficult for these objects, but 
\citet{Guedel_Proxima_Cen_2} and \citet{Fuhrmeister_CN_Leo} found the inverse FIP effect to be less pronounced in the M5.5 stars Proxima~Cen and CN~Leo. With our systematic line ratio survey, we can consolidate the trend of the inverse FIP effect in M~dwarf coronae weakening with later spectral subclass. This is coupled to the decrease of the activity observed with later subclass in general, while $L_{\mathrm{X}}/L_{\mathrm{bol}}$ is similar and dominated by saturation effects for all our sample stars except for the latest object Proxima~Cen. 
Apart from the precise strength of single emission lines that are mostly determined by individual abundances, high resolution
X-ray spectra of M~dwarfs have a very similar overall shape. Only the continuum level decreases with later spectral type, as lower coronal temperatures reduce the amount of true continuum emission as well as the pseudo-continuum dominated by lines of highly ionized iron located at approximately 9--17~\AA. The reduction of the continuum level can easily be quantified from the cumulative spectra and serves as an indicator for the average coronal temperature.

Both methods we used for our analysis, the cumulative spectra as well as the emission measure independent line ratios, can be applied to spectra with lower signal where only the strongest lines can be discerned. They are thus also suitable to determine coronal properties from the spectra of the fainter mid and late M~dwarfs. Only very few grating spectra of such objects are available so far, and especially spectra of stars later than M6 would help to confirm the trend that the difference of the Ne/Fe and Mg/Ne ratios to solar photospheric values decreases with later spectral type. However, since the Ne/O and Si/Mg ratios stay enhanced by about a factor of two, there is no indication of all abundance anomalies to diminish. So at least in comparison to the saturated coronae of active M dwarfs, it is the solar photosphere which appears anomalous; and the abundances of M~dwarfs seem to ``converge'' with later spectral subclass, but to values differing from the solar photospheric level.

The determination of reference sets of photospheric abundances of active M dwarfs is rather difficult, since reliable models of the stellar atmosphere including radiative transfer are required. The parameters involved in the modeling procedure, i.\,e. the stellar mass, effective temperature, $\log g$, and for M~dwarfs, the influence of the stellar chromosphere and possibly the formation of dust, introduce large uncertainties in the derived abundances \citep{mdwarf_photospheric}. Additionally, only the abundances of heavier elements, often only that of iron, can be determined in the photosphere, while light elements like carbon, nitrogen and oxygen or noble gases like neon that dominate the coronal spectrum leave only weak or no signatures in the photospheric spectrum. 

It may be questionable to line up active M~dwarfs exclusively based on spectral type, since other parameters, especially the age, certainly have a non-negligible influence, and age-activity relations are well-established over all wavelengths and in particular for coronal X-ray emission \citep{Vaiana_X-ray_age, Favata_X-ray_age}. Our ultraluminous sample star AU~Mic forms a proper motion pair with the M4/M4.5 binary \object{AT Mic}, both stars are members of the $\beta$~Pictoris moving group and only about 10--12 Myrs old \citep{Zuckerman_beta_Pic_group}. \citet{Robrade_M_dwarfs} found AT~Mic half an order of magnitude brighter in X-rays than our sample stars AD~Leo, EV~Lac and EQ~Peg, which have similar spectral types. Stars with a common origin, like the two ZAMS stars AU~Mic and AT~Mic, or the co-eval components of the binary EQ~Peg thus provide a testbed independent of such biases. 
The M3.5 and M4.5 components of EQ~Peg differ by one spectral subclass, but the X-ray luminosity of the M4.5 secondary is considerably lower, which we can attribute to a fairly reduced amount of coronal emission measure, but at temperatures similar to those found on the primary. Yet, in terms of $L_{\mathrm{X}}/L_{\mathrm{bol}}$,  both stars are in the saturation regime, and thus again comparable. Similar considerations apply to a larger sample of different ages, ranging from a few Myrs (AU~Mic) to almost one Gyr (EQ~Peg); the cumulative spectra of very young stars like AU~Mic or EV~Lac are very similar to the much older EQ~Peg system. Proxima Cen is also by far the oldest object, and the one with the longest rotation period. It is the only star from our sample where $L_{\mathrm{X}}/L_{\mathrm{bol}}$ deviates from the saturation level, yet it still fits the picture of the other saturated stars in terms of abundance trends. 
Here we probably see the well-established relationships between X-ray luminosity and age or, related via spin-down with age, rotation period/rotational velocity and X-ray luminosity \citep{Pizzolato_saturated, Reiners_rotation} or other activity indicators like H$\alpha$ or \ion{Ca}{ii} emission at work. The age of the other sample stars is less than 1 Gyr and their rotation periods are typically only a few days.
The sample of M~dwarfs observed with high spectral resolution in X-rays is -- by necessity --
biased towards the more luminous younger objects, and it would be crucial to examine how further old and inactive objects like e.\,g. Barnard's star fit into that picture.

During quiescence, coronal densities on active stars typically  
range between $10^{10}$--$10^{11}$~cm$^{-3}$ for the low-temperature plasma ($\approx 2$~MK) traced by the \ion{O}{vii} triplet \citep{densities_activity}, and between $10^{11}$--$10^{12}$~cm$^{-3}$ for \ion{Ne}{ix} ($\approx 4$~MK).  
Significantly higher values have so far only been observed during huge flares \citep{Guedel_Proxima_Cen_1, CN_Leo_flare}, but in general, coronal densities tend to increase during periods of increased activity and lower amplitude flaring \citep[e.\,g.][]{Maggio_Ness_AD_Leo, Fuhrmeister_CN_Leo}. \citet{Mitra_Kraev_densities} thus developed a statistical approach that showed that the different $f/i$-ratios observed during flares and quiescence on EV~Lac are actually highly significant.

Flares can considerably change the X-ray properties of active M~dwarf stars. The flare plasma often dominates the coronal emission, although the dimensions of the flaring loops seldomly reach the size of the star itself, and the flaring region typically covers only a small fraction of the stellar surface. The plasma temperature and emission measure can increase by more than an order of magnitude during the strongest flares, which subsequently results in an enhanced level of continuum emission. The bulk of line emission is dominated by high-temperature ions. Based on current flare models, the flare plasma is largely composed of evaporated chromospheric or photospheric material filling the loop. In this context, abundance gradients between the different layers of the atmosphere, like the FIP effect observed on the Sun, become important. \citet{Nordon_flares} and \citet{Nordon_flares2} investigated changes in the FIP bias during several larger flares on a sample of different active stars. For the very active stars that show an inverse FIP effect during quiescence, they found abundance ratios of the scheme high FIP/low FIP like Ne/Fe to decrease during flares, i.\,e. the inverse FIP effect turns into a solar-like pattern or even a FIP effect. This is in good agreement with the trends we find in our M dwarf sample. Thus abundance gradients, resulting in a different composition of the evaporated flare plasma, seem to be a common feature, and in contradiction to the corona, the underlying abundances of the lower atmospheric layers in active M~dwarfs approach a more solar photospheric like pattern. This finding is also supported by \citet{CN_Leo_flare}, who found the coronal iron abundance of the M5.5 dwarf \object{CN Leo} to increase by a factor of two to the solar photospheric level during a large flare. 

To what extent flares do have to be considered as the ``usual'' behavior of a stellar corona? Early M~dwarfs are in a state of permanent flaring, and even during phases of apparent quiescence, the underlying basal coronal flux is difficult to disentangle from unresolved weaker flares. While larger flares can be separated, the characteristic flickering of early M~dwarfs has to be taken as their typical ``quasi-quiescent'' behavior. The average coronal temperatures and densities of active stars will lower when the data is restricted to this quiescent state. However, we do not find that the flares included in the total datasets of our seven sample stars introduce large changes in the observed abundance ratios. Since the data do not include huge flares, this indicates a smooth transition between the quasi-quiescent state and intermediate flares, which supports the assumption that the unresolved flickering is indeed composed of smaller flares. When switching to intermediate and late M~dwarfs and their single, discrete flares, it may on the other hand be possible to pin down the true basal coronal flux as observed on the Sun during quiescence, which seems to approach solar coronal conditions for these less active stars. 
The smooth transition in X-ray luminosity, flare rate, coronal temperatures and abundances we observe in our sequence of
very active early M dwarfs towards intermediate and late M~dwarfs, i.\,e., from stars with radiative interiors to fully convective
interiors, suggests that the different dynamo mechanisms thought to operate in these stars to not lead to easily observable consequences in their coronal properties.

\section{Summary and conclusions}
We have investigated the coronal properties of the M3.5/M4.5 EQ~Peg binary system from their \emph{Chandra} HETGS spectra. No large flares occur during our observations, and we find coronal densities ranging from $4 \cdot 10^{10}$~cm$^{-3}$ to $3 \cdot 10^{11}$~cm$^{-3}$ or consistent with the low-density limit from the \ion{O}{vii} and \ion{Ne}{ix} triplets, i.\,e. values typical for active stars. The ratio of X-ray luminosities is 6:1 for EQ~Peg~A and B, but in terms of $L_{\mathrm{X}}/L_{\mathrm{bol}}$, both stars are saturated. The differential emission measures of both components peak around 3~MK, and their abundances are similar and follow the inverse FIP effect.

We compared all seven M~dwarfs observed so far with the \emph{Chandra} HETGS with two methods also suitable for spectra with low signal. The slope of the cumulative spectrum, which traces the continuum level and therefore the average coronal temperature, is very similar for spectral types  M0.5 to M4 and then decreases. Emission measure-independent abundance ratios based on the line fluxes of strong lines with a similar temperature dependence confirm the existence of abundance anomalies compared to the solar photosphere for all M~dwarfs in our sample. The ratios sensitive to a FIP bias, i.\,e. Mg/Ne and Ne/Fe, show a clear trend with increasing spectral type to approach the solar photospheric level, while ratios insensitive to a FIP bias like Si/Mg and Ne/O stay at a constant level. These trends seem to be independent of the age of the stars.

\begin{acknowledgements}
C.L. acknowledges financial support by the DLR under 50OR0105.
J.-U.N. gratefully acknowledges support provided by NASA
through Chandra Postdoctoral Fellowship grant PF5-60039
awarded by the Chandra X-ray Center, which is operated by
the Smithsonian Astrophysical Observatory for NASA under
contract NAS8-03060.
This research made use of the SIMBAD database, operated at CDS, Strasbourg, France. CHIANTI is a collaborative project involving the NRL (USA), RAL (UK), MSSL (UK), the Universities of Florence (Italy) and Cambridge (UK), and George Mason University (USA). 
\end{acknowledgements}

\bibliographystyle{aa}
\bibliography{../literature}

\begin{thebibliography}{85}
\expandafter\ifx\csname natexlab\endcsname\relax\def\natexlab#1{#1}\fi

\bibitem[{{Acton} {et~al.}(1975){Acton}, {Catura}, \&
  {Joki}}]{Acton_neon_to_oxygen}
{Acton}, L.~W., {Catura}, R.~C., \& {Joki}, E.~G. 1975, \apjl, 195, L93

\bibitem[{{Asplund} {et~al.}(2005){Asplund}, {Grevesse}, \& {Sauval}}]{Asplund}
{Asplund}, M., {Grevesse}, N., \& {Sauval}, A.~J. 2005, in ASP Conf. Ser. 336:
  Cosmic Abundances as Records of Stellar Evolution and Nucleosynthesis, 25--38

\bibitem[{{Benedict} {et~al.}(1998){Benedict}, {McArthur}, {Nelan}, {Story},
  {Whipple}, {Shelus}, {Jefferys}, {Hemenway}, {Franz}, {Wasserman},
  {Duncombe}, {van Altena}, \& {Fredrick}}]{Benedict_P_rot_Proxima}
{Benedict}, G.~F., {McArthur}, B., {Nelan}, E., {et~al.} 1998, \aj, 116, 429

\bibitem[{{Brosius} {et~al.}(1996){Brosius}, {Davila}, {Thomas}, \&
  {Monsignori-Fossi}}]{Brosius}
{Brosius}, J.~W., {Davila}, J.~M., {Thomas}, R.~J., \& {Monsignori-Fossi},
  B.~C. 1996, \apjs, 106, 143

\bibitem[{{Browning}(2008)}]{Browning_fullyconvective}
{Browning}, M.~K. 2008, \apj, 676, 1262

\bibitem[{{Chabrier} \& {Baraffe}(1997)}]{Chabrier_Baraffe}
{Chabrier}, G. \& {Baraffe}, I. 1997, \aap, 327, 1039

\bibitem[{{Chabrier} \& {K{\"u}ker}(2006)}]{Chabrier_Kuker}
{Chabrier}, G. \& {K{\"u}ker}, M. 2006, \aap, 446, 1027

\bibitem[{{Delfosse} {et~al.}(1998){Delfosse}, {Forveille}, {Perrier}, \&
  {Mayor}}]{Delfosse_mdwarfs}
{Delfosse}, X., {Forveille}, T., {Perrier}, C., \& {Mayor}, M. 1998, \aap, 331,
  581

\bibitem[{{Dobler} {et~al.}(2006){Dobler}, {Stix}, \&
  {Brandenburg}}]{Dobler_fullyconvective}
{Dobler}, W., {Stix}, M., \& {Brandenburg}, A. 2006, \apj, 638, 336

\bibitem[{{Drake} \& {Testa}(2005)}]{Drake_Testa}
{Drake}, J.~J. \& {Testa}, P. 2005, \nat, 436, 525

\bibitem[{{Durney} {et~al.}(1993){Durney}, {De Young}, \&
  {Roxburgh}}]{Durney_dynamo}
{Durney}, B.~R., {De Young}, D.~S., \& {Roxburgh}, I.~W. 1993, \solphys, 145,
  207

\bibitem[{{Favata} {et~al.}(1994){Favata}, {Micela}, {Sciortino}, \&
  {Barbera}}]{Favata_X-ray_age}
{Favata}, F., {Micela}, G., {Sciortino}, S., \& {Barbera}, M. 1994, in
  Astrophysics and Space Science Library, Vol. 187, Frontiers of Space and
  Ground-Based Astronomy, ed. W.~{Wamsteker}, M.~S. {Longair}, \& Y.~{Kondo},
  589--+

\bibitem[{{Favata} {et~al.}(2000){Favata}, {Reale}, {Micela}, {Sciortino},
  {Maggio}, \& {Matsumoto}}]{Favata_EV_Lac}
{Favata}, F., {Reale}, F., {Micela}, G., {et~al.} 2000, \aap, 353, 987

\bibitem[{{Feldman} \& {Laming}(2000)}]{Feldman_Laming}
{Feldman}, U. \& {Laming}, J.~M. 2000, \physscr, 61, 222

\bibitem[{{Fleming} {et~al.}(1993){Fleming}, {Giampapa}, {Schmitt}, \&
  {Bookbinder}}]{Fleming_M_dwarfs}
{Fleming}, T.~A., {Giampapa}, M.~S., {Schmitt}, J.~H.~M.~M., \& {Bookbinder},
  J.~A. 1993, \apj, 410, 387

\bibitem[{{Fuhrmeister} {et~al.}(2007){Fuhrmeister}, {Liefke}, \&
  {Schmitt}}]{Fuhrmeister_CN_Leo}
{Fuhrmeister}, B., {Liefke}, C., \& {Schmitt}, J.~H.~M.~M. 2007, \aap, 468, 221

\bibitem[{{G{\"u}del} {et~al.}(2001){G{\"u}del}, {Audard}, {Magee},
  {Franciosini}, {Grosso}, {Cordova}, {Pallavicini}, \& {Mewe}}]{Guedel_YY_Gem}
{G{\"u}del}, M., {Audard}, M., {Magee}, H., {et~al.} 2001, \aap, 365, L344

\bibitem[{{G{\"u}del} {et~al.}(2004){G{\"u}del}, {Audard}, {Reale}, {Skinner},
  \& {Linsky}}]{Guedel_Proxima_Cen_2}
{G{\"u}del}, M., {Audard}, M., {Reale}, F., {Skinner}, S.~L., \& {Linsky},
  J.~L. 2004, \aap, 416, 713

\bibitem[{{G{\"u}del} {et~al.}(2002{\natexlab{a}}){G{\"u}del}, {Audard},
  {Skinner}, \& {Horvath}}]{Guedel_Proxima_Cen_1}
{G{\"u}del}, M., {Audard}, M., {Skinner}, S.~L., \& {Horvath}, M.~I.
  2002{\natexlab{a}}, \apjl, 580, L73

\bibitem[{{G{\"u}del} {et~al.}(2002{\natexlab{b}}){G{\"u}del}, {Audard},
  {Sres}, {Wehrli}, {Behar}, {Mewe}, {Raassen}, \& {Magee}}]{Guedel_FIP_IFIP}
{G{\"u}del}, M., {Audard}, M., {Sres}, A., {et~al.} 2002{\natexlab{b}}, in ASP
  Conf. Ser. 277: Stellar Coronae in the Chandra and XMM-NEWTON Era, ed.
  F.~{Favata} \& J.~J. {Drake}, 497--501

\bibitem[{{Haisch} {et~al.}(1987){Haisch}, {Butler}, {Doyle}, \&
  {Rodono}}]{Haisch_EQ_Peg}
{Haisch}, B.~M., {Butler}, C.~J., {Doyle}, J.~G., \& {Rodono}, M. 1987, \aap,
  181, 96

\bibitem[{{Hawley} {et~al.}(2003){Hawley}, {Allred}, {Johns-Krull}, {Fisher},
  {Abbett}, {Alekseev}, {Avgoloupis}, {Deustua}, {Gunn}, {Seiradakis}, {Sirk},
  \& {Valenti}}]{Hawley_Allred}
{Hawley}, S.~L., {Allred}, J.~C., {Johns-Krull}, C.~M., {et~al.} 2003, \apj,
  597, 535

\bibitem[{{Jackson} {et~al.}(1989){Jackson}, {Kundu}, \&
  {White}}]{Jackson_radio}
{Jackson}, P.~D., {Kundu}, M.~R., \& {White}, S.~M. 1989, \aap, 210, 284

\bibitem[{{Katsova} {et~al.}(2002){Katsova}, {Livshits}, \&
  {Schmitt}}]{Katsova_EQ_Peg}
{Katsova}, M.~M., {Livshits}, M.~A., \& {Schmitt}, J.~H.~M.~M. 2002, in
  Astronomical Society of the Pacific Conference Series, Vol. 277, Stellar
  Coronae in the Chandra and XMM-NEWTON Era, ed. F.~{Favata} \& J.~J. {Drake},
  515--520

\bibitem[{{Kenyon} \& {Hartmann}(1995)}]{Kenyon_Hartmann}
{Kenyon}, S.~J. \& {Hartmann}, L. 1995, \apjs, 101, 117

\bibitem[{{Kron}(1952)}]{Kron_P_rot_YY_Gem}
{Kron}, G.~E. 1952, \apj, 115, 301

\bibitem[{{Lacy} {et~al.}(1976){Lacy}, {Moffett}, \& {Evans}}]{Lacy_flarestars}
{Lacy}, C.~H., {Moffett}, T.~J., \& {Evans}, D.~S. 1976, \apjs, 30, 85

\bibitem[{{Landi} {et~al.}(2006){Landi}, {Del Zanna}, {Young}, {Dere}, {Mason},
  \& {Landini}}]{Chianti7}
{Landi}, E., {Del Zanna}, G., {Young}, P.~R., {et~al.} 2006, \apjs, 162, 261

\bibitem[{{Liefke} {et~al.}(in preparation){Liefke}, {Fuhrmeister}, \&
  {Schmitt}}]{CN_Leo_flare}
{Liefke}, C., {Fuhrmeister}, B., \& {Schmitt}, J.~H.~M.~M. in preparation

\bibitem[{{Liefke} {et~al.}(2007){Liefke}, {Reiners}, \&
  {Schmitt}}]{Coronae_Proceedings}
{Liefke}, C., {Reiners}, A., \& {Schmitt}, J.~H.~M.~M. 2007, Memorie della
  Societa Astronomica Italiana, 78, 258

\bibitem[{{Liefke} \& {Schmitt}(2006)}]{alpha_Cen_Ne/O}
{Liefke}, C. \& {Schmitt}, J.~H.~M.~M. 2006, \aap, 458, L1

\bibitem[{{Liefke} \& {Schmitt}(in preparation)}]{DEM}
{Liefke}, C. \& {Schmitt}, J.~H.~M.~M. in preparation

\bibitem[{{Linsky} {et~al.}(2002){Linsky}, {Ayres}, {Brown}, \&
  {Osten}}]{Linsky_AU_Mic}
{Linsky}, J.~L., {Ayres}, T.~R., {Brown}, A., \& {Osten}, R.~A. 2002, Astron.
  Nachr., 323, 321

\bibitem[{{Magee} {et~al.}(2003){Magee}, {G{\"u}del}, {Audard}, \&
  {Mewe}}]{Magee_AU_Mic}
{Magee}, H.~R.~M., {G{\"u}del}, M., {Audard}, M., \& {Mewe}, R. 2003, Advances
  in Space Research, 32, 1149

\bibitem[{{Maggio} {et~al.}(2004){Maggio}, {Drake}, {Kashyap}, {Harnden},
  {Micela}, {Peres}, \& {Sciortino}}]{Maggio_AD_Leo}
{Maggio}, A., {Drake}, J.~J., {Kashyap}, V., {et~al.} 2004, \apj, 613, 548

\bibitem[{{Maggio} \& {Ness}(2005)}]{Maggio_Ness_AD_Leo}
{Maggio}, A. \& {Ness}, J.-U. 2005, \apjl, 622, L57

\bibitem[{{Mahmoud}(1993)}]{Mahmoud_P_rot_AD_Leo}
{Mahmoud}, F.~M. 1993, \apss, 209, 237

\bibitem[{{Mazzotta} {et~al.}(1998){Mazzotta}, {Mazzitelli}, {Colafrancesco},
  \& {Vittorio}}]{Mazzotta_etal}
{Mazzotta}, P., {Mazzitelli}, G., {Colafrancesco}, S., \& {Vittorio}, N. 1998,
  \aaps, 133, 403

\bibitem[{{Mitra-Kraev} \& {Ness}(2006)}]{Mitra_Kraev_densities}
{Mitra-Kraev}, U. \& {Ness}, J.-U. 2006, in High Resolution X-ray Spectroscopy:
  towards XEUS and Con-X

\bibitem[{{Moffatt}(1978)}]{Moffat_dynamo}
{Moffatt}, H.~K. 1978, {Magnetic field generation in electrically conducting
  fluids} (Cambridge, England, Cambridge University Press, 1978.~353 p.)

\bibitem[{{Monsignori Fossi} {et~al.}(1995){Monsignori Fossi}, {Landini},
  {Fruscione}, \& {Dupuis}}]{EQ_Peg_EUVE}
{Monsignori Fossi}, B.~C., {Landini}, M., {Fruscione}, A., \& {Dupuis}, J.
  1995, \apj, 449, 376

\bibitem[{{Mullan} \& {MacDonald}(2001)}]{Mullan_fullyconvective}
{Mullan}, D.~J. \& {MacDonald}, J. 2001, \apj, 559, 353

\bibitem[{{Ness} {et~al.}(2003{\natexlab{a}}){Ness}, {Audard}, {Schmitt}, \&
  {G{\" u}del}}]{densities_activity}
{Ness}, J.-U., {Audard}, M., {Schmitt}, J.~H.~M.~M., \& {G{\" u}del}, M.
  2003{\natexlab{a}}, Advances in Space Research, 32, 937

\bibitem[{{Ness} {et~al.}(2003{\natexlab{b}}){Ness}, {Brickhouse}, {Drake}, \&
  {Huenemoerder}}]{Capella_Ne}
{Ness}, J.-U., {Brickhouse}, N.~S., {Drake}, J.~J., \& {Huenemoerder}, D.~P.
  2003{\natexlab{b}}, \apj, 598, 1277

\bibitem[{{Ness} {et~al.}(2004){Ness}, {G{\"u}del}, {Schmitt}, {Audard}, \&
  {Telleschi}}]{sizes}
{Ness}, J.-U., {G{\"u}del}, M., {Schmitt}, J.~H.~M.~M., {Audard}, M., \&
  {Telleschi}, A. 2004, \aap, 427, 667

\bibitem[{{Ness} \& {Jordan}(2008)}]{ness_jordan}
{Ness}, J.-U. \& {Jordan}, C. 2008, \mnras, 385, 1691

\bibitem[{{Ness} {et~al.}(2003{\natexlab{c}}){Ness}, {Schmitt}, {Audard}, {G{\"
  u}del}, \& {Mewe}}]{opacity}
{Ness}, J.-U., {Schmitt}, J.~H.~M.~M., {Audard}, M., {G{\" u}del}, M., \&
  {Mewe}, R. 2003{\natexlab{c}}, \aap, 407, 347

\bibitem[{{Ness} {et~al.}(2002){Ness}, {Schmitt}, {Burwitz}, {Mewe}, \&
  {Predehl}}]{Algol_LETGS}
{Ness}, J.-U., {Schmitt}, J.~H.~M.~M., {Burwitz}, V., {Mewe}, R., \& {Predehl},
  P. 2002, \aap, 387, 1032

\bibitem[{{Ness} \& {Wichmann}(2002)}]{CORA}
{Ness}, J.-U. \& {Wichmann}, R. 2002, Astron. Nachr., 323, 129

\bibitem[{{Nordon} \& {Behar}(2007)}]{Nordon_flares}
{Nordon}, R. \& {Behar}, E. 2007, \aap, 464, 309

\bibitem[{{Nordon} \& {Behar}(2008)}]{Nordon_flares2}
{Nordon}, R. \& {Behar}, E. 2008, \aap, 482, 639

\bibitem[{{Norton} {et~al.}(2007){Norton}, {Wheatley}, {West}, {Haswell},
  {Street}, {Collier Cameron}, {Christian}, {Clarkson}, {Enoch}, {Gallaway},
  {Hellier}, {Horne}, {Irwin}, {Kane}, {Lister}, {Nicholas}, {Parley},
  {Pollacco}, {Ryans}, {Skillen}, \& {Wilson}}]{Norton_P_rot_EQ_Peg}
{Norton}, A.~J., {Wheatley}, P.~J., {West}, R.~G., {et~al.} 2007, \aap, 467,
  785

\bibitem[{{Osten} {et~al.}(2006){Osten}, {Hawley}, {Allred}, {Johns-Krull},
  {Brown}, \& {Harper}}]{Osten_EV_Lac_quiescence}
{Osten}, R.~A., {Hawley}, S.~L., {Allred}, J., {et~al.} 2006, \apj, 647, 1349

\bibitem[{{Osten} {et~al.}(2005){Osten}, {Hawley}, {Allred}, {Johns-Krull}, \&
  {Roark}}]{Osten_EV_Lac_flare}
{Osten}, R.~A., {Hawley}, S.~L., {Allred}, J.~C., {Johns-Krull}, C.~M., \&
  {Roark}, C. 2005, \apj, 621, 398

\bibitem[{{Pallavicini} {et~al.}(1986){Pallavicini}, {Kundu}, \&
  {Jackson}}]{Pallavicini_EQ_Peg_EXOSAT}
{Pallavicini}, R., {Kundu}, M.~R., \& {Jackson}, P.~D. 1986, in Lecture Notes
  in Physics, Berlin Springer Verlag, Vol. 254, Cool Stars, Stellar Systems and
  the Sun, ed. M.~{Zeilik} \& D.~M. {Gibson}, 225--228

\bibitem[{{Pallavicini} {et~al.}(1990){Pallavicini}, {Tagliaferri}, \&
  {Stella}}]{Pallavicini_EXOSAT}
{Pallavicini}, R., {Tagliaferri}, G., \& {Stella}, L. 1990, \aap, 228, 403

\bibitem[{{Parker}(1955)}]{Parker_dynamo}
{Parker}, E.~N. 1955, \apj, 122, 293

\bibitem[{{Parkinson}(1977)}]{Parkinson_abundance_ratios}
{Parkinson}, J.~H. 1977, \aap, 57, 185

\bibitem[{{Parsamyan}(1995)}]{Parsamyan_ages}
{Parsamyan}, E.~S. 1995, Astrophysics, 38, 206

\bibitem[{{Pettersen}(1976)}]{Pettersen_flarestars}
{Pettersen}, B.~R. 1976, Institute of Theoretical Astrophysics Blindern Oslo
  Reports, 46, 1

\bibitem[{{Pettersen} {et~al.}(1992){Pettersen}, {Olah}, \&
  {Sandmann}}]{Pettersen_P_rot_EV_Lac}
{Pettersen}, B.~R., {Olah}, K., \& {Sandmann}, W.~H. 1992, \aaps, 96, 497

\bibitem[{{Pizzolato} {et~al.}(2003){Pizzolato}, {Maggio}, {Micela},
  {Sciortino}, \& {Ventura}}]{Pizzolato_saturated}
{Pizzolato}, N., {Maggio}, A., {Micela}, G., {Sciortino}, S., \& {Ventura}, P.
  2003, \aap, 397, 147

\bibitem[{{Raassen} {et~al.}(2003){Raassen}, {Ness}, {Mewe}, {van der Meer},
  {Burwitz}, \& {Kaastra}}]{Raassen}
{Raassen}, A.~J.~J., {Ness}, J.-U., {Mewe}, R., {et~al.} 2003, \aap, 400, 671

\bibitem[{{Reiners}(2007)}]{Reiners_rotation}
{Reiners}, A. 2007, \aap, 467, 259

\bibitem[{{Robrade} {et~al.}(2004){Robrade}, {Ness}, \&
  {Schmitt}}]{Robrade_EQ_Peg}
{Robrade}, J., {Ness}, J.-U., \& {Schmitt}, J.~H.~M.~M. 2004, \aap, 413, 317

\bibitem[{{Robrade} \& {Schmitt}(2005)}]{Robrade_M_dwarfs}
{Robrade}, J. \& {Schmitt}, J.~H.~M.~M. 2005, \aap, 435, 1073

\bibitem[{{Rutledge} {et~al.}(2000){Rutledge}, {Basri}, {Mart{\'{\i}}n}, \&
  {Bildsten}}]{Rutledge_LP944-20}
{Rutledge}, R.~E., {Basri}, G., {Mart{\'{\i}}n}, E.~L., \& {Bildsten}, L. 2000,
  \apjl, 538, L141

\bibitem[{{Sanz-Forcada} {et~al.}(2004){Sanz-Forcada}, {Favata}, \&
  {Micela}}]{coronal_photospheric}
{Sanz-Forcada}, J., {Favata}, F., \& {Micela}, G. 2004, \aap, 416, 281

\bibitem[{{Schmitt} \& {Liefke}(2002)}]{LHS_2065}
{Schmitt}, J.~H.~M.~M. \& {Liefke}, C. 2002, \aap, 382, L9

\bibitem[{{Schmitt} \& {Ness}(2002)}]{CNO}
{Schmitt}, J.~H.~M.~M. \& {Ness}, J.-U. 2002, \aap, 388, L13

\bibitem[{{Schmitt} \& {Ness}(2004)}]{Algol_EMD}
{Schmitt}, J.~H.~M.~M. \& {Ness}, J.-U. 2004, \aap, 415, 1099

\bibitem[{{Schmitt} {et~al.}(2008){Schmitt}, {Reale}, {Liefke}, {Wolter},
  {Fuhrmeister}, {Reiners}, \& {Peres}}]{CN_Leo_onset}
{Schmitt}, J.~H.~M.~M., {Reale}, F., {Liefke}, C., {et~al.} 2008, \aap, 481,
  799

\bibitem[{{Spiesman} \& {Hawley}(1986)}]{Spiesman_P_rot_AD_Leo}
{Spiesman}, W.~J. \& {Hawley}, S.~L. 1986, \aj, 92, 664

\bibitem[{{Stelzer} {et~al.}(2002){Stelzer}, {Burwitz}, {Audard}, {G{\"u}del},
  {Ness}, {Grosso}, {Neuh{\"a}user}, {Schmitt}, {Predehl}, \&
  {Aschenbach}}]{Stelzer_YY_Gem}
{Stelzer}, B., {Burwitz}, V., {Audard}, M., {et~al.} 2002, \aap, 392, 585

\bibitem[{{Stelzer} {et~al.}(2006){Stelzer}, {Schmitt}, {Micela}, \&
  {Liefke}}]{Stelzer_LP412-31}
{Stelzer}, B., {Schmitt}, J.~H.~M.~M., {Micela}, G., \& {Liefke}, C. 2006,
  \aap, 460, L35

\bibitem[{{Telleschi} {et~al.}(2005){Telleschi}, {G{\" u}del}, {Briggs},
  {Audard}, {Ness}, \& {Skinner}}]{Sun_in_time}
{Telleschi}, A., {G{\" u}del}, M., {Briggs}, K., {et~al.} 2005, \apj, 622, 653

\bibitem[{{Testa} {et~al.}(2004){Testa}, {Drake}, \& {Peres}}]{HETG_density}
{Testa}, P., {Drake}, J.~J., \& {Peres}, G. 2004, \apj, 617, 508

\bibitem[{{Testa} {et~al.}(2007){Testa}, {Drake}, {Peres}, \&
  {Huenemoerder}}]{Testa_optical_depth}
{Testa}, P., {Drake}, J.~J., {Peres}, G., \& {Huenemoerder}, D.~P. 2007, \apj,
  665, 1349

\bibitem[{{Torres} {et~al.}(1972){Torres}, {Ferraz Mello}, \&
  {Quast}}]{Torres_P_rot_AU_Mic}
{Torres}, C.~A.~O., {Ferraz Mello}, S., \& {Quast}, G.~R. 1972, \aplett, 11, 13

\bibitem[{{Torres} \& {Ribas}(2002)}]{Torres_YY_gem}
{Torres}, G. \& {Ribas}, I. 2002, The Astrophysical Journal, 567, 1140

\bibitem[{{Vaiana} {et~al.}(1992){Vaiana}, {Maggio}, {Micela}, \&
  {Sciortino}}]{Vaiana_X-ray_age}
{Vaiana}, G.~S., {Maggio}, A., {Micela}, G., \& {Sciortino}, S. 1992, Memorie
  della Societa Astronomica Italiana, 63, 545

\bibitem[{{van den Besselaar} {et~al.}(2003){van den Besselaar}, {Raassen},
  {Mewe}, {van der Meer}, {G{\"u}del}, \& {Audard}}]{vandenBesselaar_AD_Leo}
{van den Besselaar}, E.~J.~M., {Raassen}, A.~J.~J., {Mewe}, R., {et~al.} 2003,
  \aap, 411, 587

\bibitem[{{Y{\i}ld{\i}z}(2007)}]{Yildiz_alpha_Cen}
{Y{\i}ld{\i}z}, M. 2007, \mnras, 374, 1264

\bibitem[{{Zboril} \& {Byrne}(1998)}]{mdwarf_photospheric}
{Zboril}, M. \& {Byrne}, P.~B. 1998, \mnras, 299, 753

\bibitem[{{Zuckerman} {et~al.}(2001){Zuckerman}, {Song}, {Bessell}, \&
  {Webb}}]{Zuckerman_beta_Pic_group}
{Zuckerman}, B., {Song}, I., {Bessell}, M.~S., \& {Webb}, R.~A. 2001, \apjl,
  562, L87

\end{thebibliography}

\end{document}